\newif\ifdraft
\draftfalse

\newif\ifTR
\TRtrue

\ifdraft
\documentclass[10pt,journal,compsoc]{IEEEtran}
\else
\documentclass[10pt,journal,compsoc]{IEEEtran}
\fi

\ifdraft
  \usepackage{drafts}
\else
  \usepackage[nodrafts]{drafts}
\fi

\usepackage{amsthm}
\usepackage{amssymb}
\usepackage{textcomp}
\usepackage{amsmath}

\usepackage[T1]{fontenc}
\usepackage[twocolumn]{figures}

\usepackage{mdframed}
\usepackage{threeparttable}
\usepackage{booktabs}
\usepackage{makecell}
\usepackage{xspace}
\usepackage{balance}
\usepackage{microtype}
\usepackage[inline]{enumitem}
\usepackage{xcolor}
\usepackage{colortbl}
\usepackage{listings}
\usepackage{url}
\usepackage{algorithm} 
\usepackage{algpseudocode} 
\algnewcommand{\LineComment}[1]{\Statex \hskip\algorithmicindent \(\triangleright\) #1}
\algrenewcommand\algorithmicindent{1em}%
\usepackage{tikz}
\usepackage{pgfplots}   
\usepackage{filecontents}
\usepackage{subcaption}
\usepackage{wrapfig}

\theoremstyle{definition}
\newtheorem{definition}{Definition}
\newtheorem{theorem}{Theorem}
\newtheorem{corollary}{Corollary}

\algnewcommand\algorithmicforeach{\textbf{for each}}
\algdef{S}[FOR]{ForEach}[1]{\algorithmicforeach\ #1\ \algorithmicdo}

\newcommand\PLOTSUBFIG{0.23\textwidth}
\newcommand\PLOTWIDTH{0.45}
\newcommand\MDEOPLOTCOLOR{blue}
\newcommand\MODAGAMEPLOTCOLOR{green}
\newcommand\SATIBEAPLOTCOLOR{red}
\newcommand\MDEOPLOTSTYLE{solid}
\newcommand\MODAGAMEPLOTSTYLE{dashed}
\newcommand\SATIBEAPLOTSTYLE{dotted}

\newcommand\PLOTGENERALSTYLE{ultra thick}

\newcommand\hlcell{{\cellcolor[gray]{.8}}}

\newcommand{\mypar}[1]{\smallskip
\noindent{}\textbf{#1.}}
\newcommand{\mysubpar}[1]{\smallskip \noindent{}\textit{#1.}}
\newcommand{\mysubparproof}[1]{\noindent{}\textit{#1}:}

\def\ie{{\emph{i.e.,}}}
\def\eg{{\emph{e.g.,}}}

\def\aka{{\emph{aka},}}

\def\wrt{{\emph{wrt.}}}

\def\st{{\emph{s.t.}}}

\def\ACAPULCO{aCaPulCO}
\def\SATIBEA{SATIBEA}
\def\MODAGAME{MODAGAME}

\ifCLASSOPTIONcompsoc
  \usepackage[nocompress]{cite}
\else
  \usepackage{cite}
\fi

\ifCLASSINFOpdf
\else
\fi

\begin{document}

\title{\emph{We're Not Gonna Break It!} Consistency-Preserving Operators for Efficient Product Line Configuration}

\author{Jose-Miguel~Horcas, Daniel~Str\"uber, Alexandru~Burdusel, Jabier Martinez, and Steffen Zschaler
\IEEEcompsocitemizethanks{
\IEEEcompsocthanksitem J. M. Horcas is with the CAOSD Group, ITIS, Universidad de M\'alaga, Spain. E-mail: horcas@lcc.uma.es
\IEEEcompsocthanksitem D. Str\"uber is with Chalmers | University Gothenburg, Sweden, and Radboud University Nijmegen, 
The Netherlands.
\IEEEcompsocthanksitem A. Burdusel is with the Department of Informatics, King's College London, United Kingdom.
\IEEEcompsocthanksitem J. Martinez is with Tecnalia, Basque Research and Technology Alliance (BRTA), Derio, Spain.
\IEEEcompsocthanksitem S. Zschaler  is with the Department of Informatics, King's College London, United Kingdom.
}
}

%
%

\markboth{IEEE Transactions on Software Engineering}%
{Horcas \MakeLowercase{\textit{et al.}}: Bare Demo of IEEEtran.cls for Computer Society Journals}
%



\IEEEtitleabstractindextext{%
\begin{abstract}
  When configuring a software product line, finding a good trade-off between multiple orthogonal quality concerns is a challenging multi-objective optimisation problem.
  State-of-the-art solutions based on search-based techniques create invalid configurations in intermediate steps, requiring additional repair actions that reduce the efficiency of the search.
  In this work, we introduce \textsl{consistency-preserving configuration operators} (CPCOs)---genetic operators that maintain valid configurations throughout the entire search.
	CPCOs bundle coherent sets of changes: the activation or deactivation of a particular feature together with other (de)activations that are needed to preserve validity. 
	In our evaluation, our instantiation of the IBEA algorithm with CPCOs outperforms two state-of-the-art tools for optimal product line configuration in terms of both speed and solution quality.
	The improvements are especially pronounced in large product lines with thousands of features.
\end{abstract}

}

\maketitle

\IEEEdisplaynontitleabstractindextext

%
\IEEEpeerreviewmaketitle


%
%
%
%
\IEEEraisesectionheading{\section{Introduction}\label{section:introduction}}

\IEEEPARstart{S}{oftware} Product Lines (SPL) aim to capture a range of related software products by dividing them into features and giving an explicit model of how features can be combined to form valid products \cite{featureorientedSPL}. Such a \emph{feature model}~\cite{foda} captures features as well as constraints about which features must be part of every product, which features are mutually exclusive, or which features require other features to also be part of the same product. A feature model, thus, can easily describe a very large set of products. \emph{Extended feature models}~\cite{benavides2005automated}, which annotate features with additional quantitative information, ask for configurations that are not only valid, but also optimal with regard to a set of non-functional attributes,
for example, maximizing performance and minimizing energy consumption.
Since each feature contributes to multiple, orthogonal objectives, the resulting search spaces are complex with many local optima, rendering the problem NP-complete \cite{literatureReview}.
  
  Optimization for configurable software is a widely-studied activity \cite{pereira2021slr} and automated optimal configuration has been studied for 15 years now \cite{benavides2005automated}.
Recent experimental evaluations show that the most scalable approaches are search-based ones that rely on metaheuristic algorithms, specifically, the genetic algorithm IBEA \cite{sayyad2013scalable,sayyad2013value,modagame,siegmund2017attributed,GuoICSE18}.
Solutions typically  encode feature selection information into a binary genotype and use standard genetic operators for mutation and crossover~\cite{literatureReview,sayyad2013scalable,sayyad2013value}. 
%
%
They also support multi-objective optimisation, which is useful for supporting extended feature models with multiple orthogonal objectives. 

The standard genetic operators currently used tend to create invalid configurations, which need to be repaired during the search.
For example, \MODAGAME~\cite{modagame} includes a dedicated ``fix'' operator, and \SATIBEA~\cite{satibea} introduces an additional mutation operator for repairing violations.
This has two main drawbacks:
First, computing a repair action during the search can be costly and impair the search performance.
Second, when the repair is not linked to a particular change that introduced the violation, it becomes somewhat arbitrary, thus potentially steering the search into some non-optimal direction.


In this paper, we explore the idea that significant improvements can be achieved if the search \textit{never produces invalid solutions by design}, removing the need for any form of repair during the search.
No previous work has investigated this: it requires a method for ensuring validity when applying the genetic operators, which was  previously not available.

We propose the concept of \textit{consistency-preserving configuration operators} (CPCOs)  for automated optimal configuration. A CPCO is a mutation operator that bundles coherent sets of changes, specifically, the activation or deactivation of a particular feature with other changes that are needed to preserve validity. 
CPCOs address the drawbacks of repair-based approaches as follows:
First, the need to compute repair steps during search runs is removed; the operator suite is generated ``offline'' before the search.
Second, since CPCOs encode minimal sets of changes required to preserve validity, they allow to explore the search space in a systematic way.
As a further benefit, we can naturally define a crossover operator, by splicing together parts of the sequences of CPCO applications that led to the two parent solutions.

  Implementing the generation of CPCOs na\"ively leads to scalability issues as it requires searching through a substantial space of possible repair actions, which grows very quickly in the size of the underlying feature model.
  We introduce an efficient algorithm for encoding CPCOs as variability-based transformation rules~\cite{struber2018variability}, which can represent multiple operators in a single compact rule.
  The algorithm and encoding allow us to build on the substantial advances made in SAT solving to help address the scalability challenge underlying CPCO generation.
	
	Our experimental evaluation shows that CPCOs  lead to highly efficient search operators for automated optimal configuration, in two dimensions:
	First, we find significant improvements in the quality of the obtained solutions compared to the state of the art.
	Second, we find improvements in the execution time of optimisation runs when they are executed with generated CPCOs.
	The offline CPCO generation adds a performance overhead, whose effect on the overall execution time depends on the application scenario:
	In ``one-shot'' scenarios where CPCOs are used only once, the overall total execution time can become higher than with conventional approaches, whereas the overhead becomes less important in dynamic scenarios.
	For example, in dynamic scenarios where the monitoring of quality attributes over time can trigger  periodic re-configurations based on new optimisation runs, CPCOs can be reused at no extra cost.

We make the following contributions:

\begin{enumerate}
  \item We introduce CPCOs, bundling coherent sets of changes that are required to preserve configuration validity.

  \item We introduce an efficient algorithm for generating CPCOs, as a na\"ive implementation of the formal procedure would not be scalable.
	
	\item We formally prove the soundness of the CPCO suite defined by the algorithm.
        In other words, CPCOs always lead to valid configuration changes. 
				
	\item We present a  new tool based on our concepts called \emph{\ACAPULCO} \cite{appendix},
        implementing the IBEA algorithm with mutation and crossover operators that apply CPCOs to a binary vector-based encoding of configurations.
        CPCOs are represented as transformation rules in the model transformation language \textit{Henshin}~\cite{struber2017henshin}. 
		
		
	\item We evaluate our technique on ten standard benchmark feature models, comparing the performance of \ACAPULCO\ against two state-of-the-art tools.
        This evaluation shows that our CPCOs support a more efficient search, producing better solutions in faster time than state-of-the-art tools.
        This observation is even more pronounced for larger feature models with thousands of features.
				Our evaluation artifacts are available online \cite{appendix}.
\end{enumerate}

To the best of our knowledge, the only work that has previously explored search-based optimal feature configuration with only valid solutions is Guo et al.~\cite{GuoICSE18}.
They explore validity preservation by adjusting the rate with which SATIBEA uses the SAT solver solution repair, including a parametrization in which every violation is repaired. 
Their results suggest that using distinct repair steps based on SAT solving does not generally improve search performance.
We speculate that this is because invalid solutions are fixed without reference to the mutation or crossover that created the solution.
As a result, it is possible that a repair may undo beneficial changes and may lead to large parts of the search space remaining unexplored.
In contrast, in this paper we explore the use of SPL-specific mutation operators (the CPCOs), which combine a mutation with the corresponding repair so that invalid solutions are never created in the first place.
CPCOs also ensure that the repair does not directly undo the intended change in feature activation.

\begin{figure}[t]
  \centering
  \includegraphics[width=\linewidth]{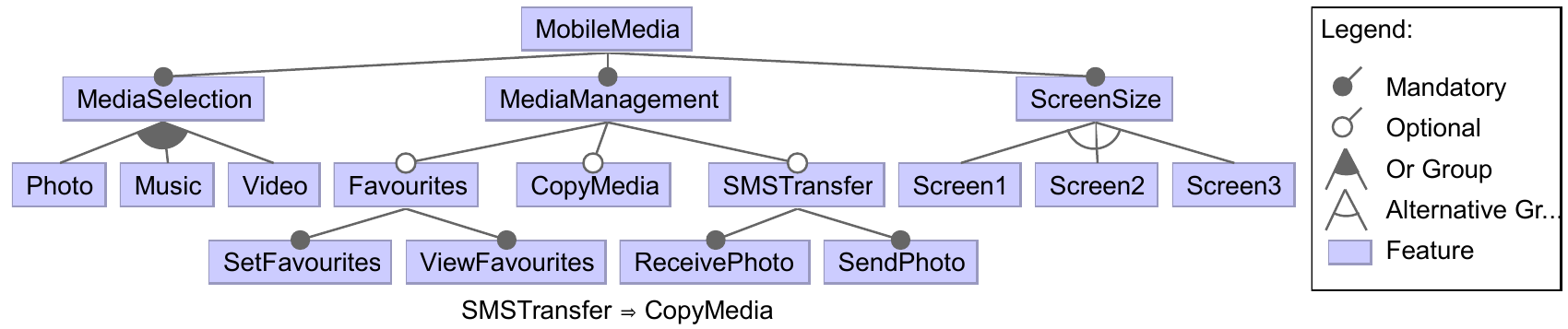}
  \caption{Example feature diagram of Mobile Media.}
  \label{fig:mobileMedia}
\end{figure}

The remainder of this paper is structured as follows: 
We first introduce CPCOs in the context of an example in Sect.~\ref{sec:example}, before describing a na\"ive algorithm for CPCO generation in Sect.~\ref{section:cpcos}.
Section~\ref{section:generating_cpcos} presents an efficient algorithm for generating CPCOs.
Section~\ref{section:evolutionary_search} explains how they can be used in the context of evolutionary search for optimal feature-model configuration, and introduces our tool that implements this idea.
Section~\ref{sec:evaluation} describes our evaluation, including a discussion of the results and their validity.
We discuss related work and conclude in Sect.~\ref{section:related_work} and \ref{section:conclusion}.
\section{CPCOs by Example}
\label{sec:example}



\mypar{Extended feature models and configurations}
Feature models~\cite{foda} allow specifying the variant space of a variant-rich system in a hierarchical form, graphically commonly represented as a feature diagram.
A feature model comprises a set of features and various relations, such as parent-to-child, alternative groups, and exclusion between features.
Each possible variant of the system arises from one configuration of the feature model, that is, a subset of its features.
The features in the subset are called \textit{active} in the given configuration.

The feature diagram in Fig.~\ref{fig:mobileMedia} represents the \textit{MobileMedia} SPL, a standard example for variability-rich systems \cite{figueiredo2008evolving}.
\textit{MobileMedia} is the root feature and has three mandatory children, of which \textit{MediaSelection} is an ``or'' group, \textit{Screensize} is an ``xor'' group, and \textit{MediaManagement} has three optional children.
The implication cross-tree constraint at the bottom represents a \textit{requires} relationship exposing the need of selecting \textit{CopyMedia} when \textit{SMSTransfer} is selected.
An example for a valid configuration is $c_{1}$ = \textit{\{MobileMedia}, \textit{MediaSelection}, \textit{Music}, \textit{MediaManagement}, \textit{ScreenSize}, \textit{Screen3\}}.
Adding \textit{Screen1} to $c_{1}$  leads to an invalid configuration, because having two active children of the same ``xor'' group violates an underlying validity  constraint of feature models.

\mypar{Optimal configuration}
In \textit{extended feature models}~\cite{benavides2005automated}, features are augmented with attributes specifying the features' contributions to non-functional attributes.
For example, consider that each feature in Fig.~\ref{fig:mobileMedia} is annotated with two attributes:

\begin{itemize}
	\item a \textit{cost}, specified as a non-negative integer value, and
	\item a \textit{benefit}, specified as a float between 0 and 10.
\end{itemize}

Optimal configuration is a multi-objective problem with several objective functions, formulated over the quality attributes (e.g., maximal benefit, minimal cost).
Solutions are \textit{Pareto fronts} of non-dominated solutions.
A solution is non-dominated if there is no other solution at least as good in all objectives and better in at least one objective.

Automated configuration is an NP-complete problem \cite{literatureReview} with complex fitness landscapes.
Available approaches to this problem~\cite{literatureReview,sayyad2013scalable,sayyad2013value} rely on conventional genetic encodings which are manipulated with standard  operators.
For example, configurations of \textit{MobileMedia} can be represented by a bit vector of length~22, where each element represents the status of a particular feature.
A standard mutation involves the toggling of a random element.
In general, toggling a random element---e.g., deactivating \textit{Screen3} in $c_1$---leads to an invalid solution that needs to be repaired.
Computing the repair can be computationally expensive and, depending on the repair strategy, lead to certain regions of the search space being neglected. 
Both factors can limit the efficiency of the search.

		
\begin{figure}[t]
  \centering
    \includegraphics[width=\linewidth]{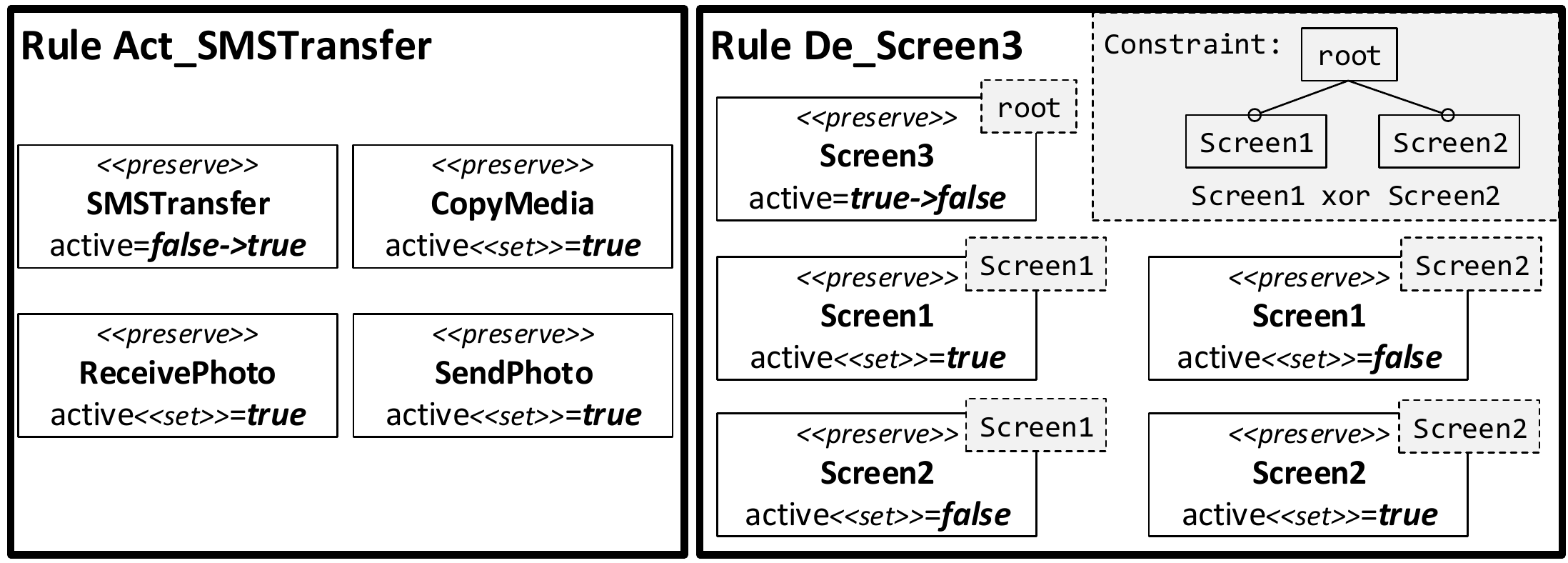}
    \caption{Example CPCOs: Activation operator for \textit{SMSTransfer}; deactivation operator for \textit{Screen3}. The latter is a \textit{variability-based rule} (VB rule) with a rule-specific feature model and presence conditions, shown with a light-grey background.}
  \label{fig:ActSMSTransfer}
\end{figure}
\begin{figure}[t]
  \centering
    \includegraphics[width=\linewidth]{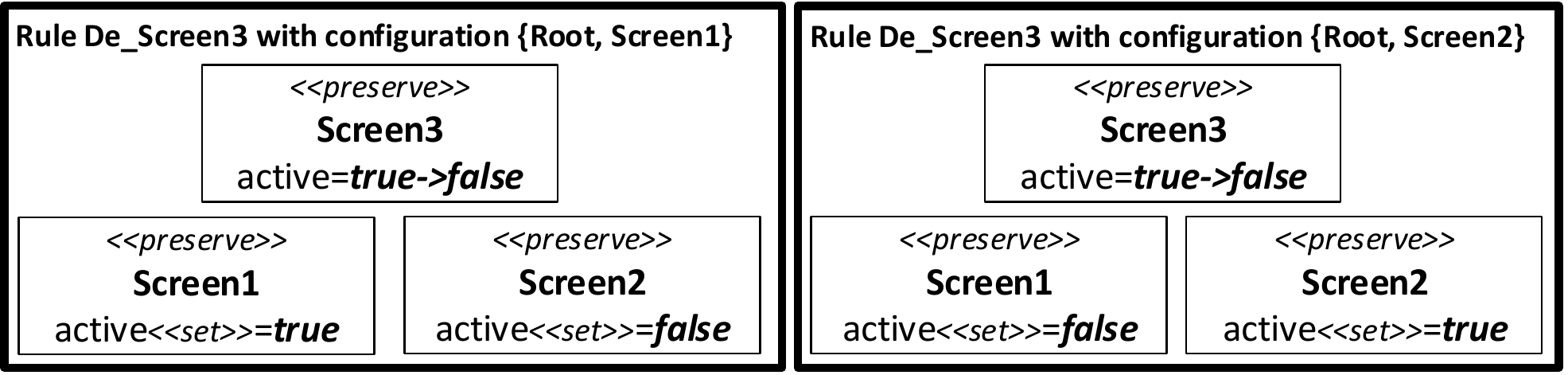}
    \caption{Two CPCO variants for the deactivation of \textit{Screen3}.}
  \label{fig:VBruleInstance}
\end{figure}


\mypar{Consistency-preserving configuration operators (CPCOs)} 
To avoid these issues, we introduce CPCOs.
These bundle coherent sets of changes:
the activation or deactivation of a particular feature and other changes that are required to retain validity.
In general, each CPCO supports multiple different sets of changes.
For example, in response to deactivating \textit{Screen3}, either \textit{Screen1} or \textit{Screen2} can be activated.
Executing a CPCO involves randomly selecting one of these change sets.
In contrast to a conventional mutation operator, each possible selection leads to a valid solution.

We define CPCOs using Henshin~\cite{struber2017henshin}, a rule-based model transformation language.
Our reason for using Henshin is twofold:
(i.) it facilitates the inspection of CPCOs in a visual syntax;
(ii.) it provides natural support for rule variants (discussed later).
In Henshin, changes to an input model (in our case, a configuration) are expressed as  graphical rules over a given metamodel (in our case, a configuration metamodel).
We generate this metamodel automatically for the feature model.
The  metamodel has a singleton meta-class for each feature with a single Boolean attribute \texttt{active} representing the activation status of the feature.



For example, Fig.~\ref{fig:ActSMSTransfer} shows two rules specifying CPCOs for \textit{MobileMedia}.
Each rule contains a set of labeled nodes and edges, where labels represent a type (\eg\ \textit{SMSTransfer}) and an action (\eg\ \textit{preserve}\footnote{In general, Henshin supports additional actions such as \textit{delete} and \textit{create} for expressing more complex transformations.}).
Nodes can have attributes, which have a type, an action (\eg\ \textit{set}), and a value.
Rule \textit{Act\_SMSTransfer} activates the feature \textit{SMSTransfer}, which necessarily includes the activation of its mandatory children \textit{ReceivePhoto} and \textit{SendPhoto}, and, due to the cross-tree constraint, feature \textit{CopyMedia}.
Therefore, the value of the  attribute \textit{active} of these features is set to \textit{true}. 
The two used attribute notations specify that the \textit{SMSTransfer} feature needs to be inactive for the rule to be applicable, whereas the other features are allowed to be already active.
 
Recall that most CPCOs encapsulate multiple possible repairs, leading to several \textit{CPCO variants}.
To avoid having many, largely redundant rules, we use a Henshin concept called \textit{variability-based (VB) rules}~\cite{struber2018variability}. 
VB rules are a compact, single-copy representation of several similar ``flat'' rules.
Elements of a VB rule are annotated with presence conditions (PCs) over a rule-specific feature model.
A PC is a propositional formula, specifying a condition under which the annotated element is present.
To derive the encoded flat rules, the feature model VB rule is \textit{configured}; that is, each feature is bound to \textit{true} or \textit{false}, and elements whose PC evaluates to \textit{false} are removed. 

Rule \textit{De\_Screen3} in Fig.~\ref{fig:ActSMSTransfer} shows an example of a VB rule that captures all the different options for repairing a deactivation of feature \textit{Screen3}.
The rule-specific feature model is shown in the top-right corner of the rule.
Note that it is significantly simpler than the original \textit{MobileMedia} feature model.
It only contains an, always active, \textit{root} feature and two optional features labelled \textit{Screen1} and \textit{Screen2}.
In addition, there is a cross-tree constraint specifying that either \textit{Screen1} or \textit{Screen2} must be selected.\footnote{The included constraint could have been expressed by an xor-group in the rule-specific feature model.
We are showing it as a separate constraint in line with the structure of VB rules our rule-generation algorithm (described in Sect.~\ref{section:generating_cpcos}) actually produces.}
The rule explicitly sets the status of \textit{Screen3} from active to inactive.
In addition, depending on the configuration of the rule-specific feature model, the rule ensures exactly one of \textit{Screen1} and \textit{Screen2} is activated instead.
Figure~\ref{fig:VBruleInstance} shows the two regular ``flat'' rules (a.k.a. CPCO variants) that can be derived from the \textit{De\_Screen3} VB rule, corresponding to the two valid configurations of the rule-specific feature model.

Note that CPCOs and their generation do not refer to the attributes from the input feature model. CPCOs are to be executed within  mutation and crossover operators of a genetic algorithm (see Sect.~\ref{section:evolutionary_search}).
The attribute values are considered by the \textit{selection} operator of that algorithm. This leads to a key advantage of CPCOs: when the attribute values change, generated CPCOs can be reused at no extra cost.


\section{Generating CPCOs: Background, principles and na\"ive generation procedure}
\label{section:cpcos}

  In this section, we explore the automated generation of consistency-preserving configuration operators (CPCOs).
  First we present background and assumptions on feature models and the considered constraints.
  Then we present and illustrate a procedure for generating CPCOs na\"ively, which does not yet scale to large feature models---an issue we address later in this paper.

  \subsection{Background and assumptions}

We start by defining the notion of feature model addressed by our CPCO generation procedure, including a feature hierarchy and cross-tree constraints.
Graphically, these concepts are typically represented in feature diagrams \cite{featureorientedSPL}.

    \newcommand{\fm}{$\textit{fm}$}
    \newcommand{\cmand}{\textsc{CMand}}
    \newcommand{\cpar}{\textsc{CPar}}
    \newcommand{\creq}{\textsc{CReq}}
    \newcommand{\cexcl}{\textsc{CExcl}}
    \newcommand{\cor}{\textsc{COr}}
    \newcommand{\cxor}{\textsc{CXor}}
    \newcommand{\croot}{\textsc{CRoot}}
    
    \begin{definition}[Feature model, Configuration]\label{def:fm}
      %

      A feature model \fm\ is a set of literals called ``features'' with a strict partial order defining a parent--child relation over the features. There exists one feature that does not have a parent and is, therefore, called the ``root'' feature.	 Child features can be mandatory or optional. Features can be group features, specifically, ``or'' and ``xor'' groups; then they must have at least two children. In addition, a feature model may define cross-tree constraints capturing further ``requires'' or ``excludes'' relationships between features.
	  
      %
      %

      Let a feature model \fm\ be given.
      A \textit{configuration} $c$ is a subset of \fm's features.
      Given a configuration $c$, a feature $f$ is \textit{active} iff $f \in c$.
      A configuration $c$ is called \textit{valid} if for all pairs of features $f, g \in \textit{fm}$, the following constraints are fulfilled:

      \begin{enumerate}
        \item \cmand: If $g$ is a mandatory child of $f$ and $f \in c$, $g \in c$.
        \item \cpar: If $g$ is the parent of $f$ and $f\in c$, $g \in c$.
        \item \creq: If $f$ requires $g$ and $f \in c$, $g \in c$.
        \item \cexcl: If $f$ excludes $g$ and $f \in c$, $g \not \in c$.
        \item \cor: If $f$ is an ``or'' group and $f \in c$, at least one of $f$'s children is active.
        \item \cxor: If $f$ is an ``xor'' group and $f \in c$, exactly one of $f$'s children is active.
        \item \croot: If $f$ is the root feature of \fm, $f \in c$.
      \end{enumerate}
      
    \end{definition}

    Fig.~\ref{fig:mobileMedia} shows an example feature model.
    The discussion at the start of Sect.~\ref{sec:example} gives examples of valid configurations for this feature model.

    A feature $f$ is either a \textit{core} feature (\ie\ every valid configuration includes~$f$), a \textit{dead} one (\ie\ no valid configuration includes~$f$), or a \textit{real-optional} one (\ie\ there exist valid configurations $c1,c2$ \st\ $f \in c1, f \notin c2$; \aka\ variant features~\cite{benavides}).
    The classification of a feature into core, real-optional or dead might not necessarily align with whether the feature is an optional child of its parent feature.
    For example, \textit{SetFavourites} is a mandatory child, but real-optional, since its parent is not mandatory.

    A feature model is called \emph{satisfiable} iff there is at least one valid configuration for it.

    We assume that the feature model is satisfiable, that it does not contain dead features, and that core and real-optional features have been computed. This can be established with available tool support~\cite{featureide} with help of  SAT solvers, which are known to be efficient for problem instances in the size of feature models in practice.
		Satisfiability analysis can be formulated as one SAT solver call per feature model.
		The other analyses lead to one SAT solver call per feature \cite{featureorientedSPL}.

  \subsection{Principle-based generation procedure}

	
  \label{sec:activate}
  
    \newcommand{\actmand}{\textsc{ActMand}}
    \newcommand{\actgroup}{\textsc{ActGroup}}
    \newcommand{\actxor}{\textsc{ActXor}}
    \newcommand{\actexcl}{\textsc{ActExc}}
    \newcommand{\actreq}{\textsc{ActReq}}
    \newcommand{\actpar}{\textsc{ActPar}}
    \newcommand{\actreqi}{\textsc{ActReq1}}
    \newcommand{\actreqii}{\textsc{ActReq2}}
    \newcommand{\dechild}{\textsc{DeChild}}
    \newcommand{\dereq}{\textsc{DeReq}}
    \newcommand{\deroot}{\textsc{DeRoot}}
    \newcommand{\deparent}{\textsc{DeParent}}
    \newcommand{\dexor}{\textsc{DeXor}}
    \newcommand{\deor}{\textsc{DeOr}}
    \newcommand{\degroup}{\textsc{DeGroup}}

 In the following, we will use the term `feature decision' to refer to an individual decision about the activation or deactivation of a specific feature.

		We present a na\"ive procedure for generating a CPCO for a given feature decision.
		To generate a full CPCO suite for a given feature model, we apply this procedure to each possible feature decision for a real-optional feature.
		That is, we generate two CPCOs for each real-optional feature: one for its activation and one for its deactivation.
			
		Our procedure is based on a notion of \textit{principles}, which are applied recursively, as long as one of them is applicable, starting from the given feature decision.
		A principle formulates a set of actions to be performed in response to a previous feature decision. Each action avoids that a constraint violation (see constraints 1--7) arises.
		Since a considered feature decision can be either an activation or deactivation, we formulate a set of activation and deactivation principles:

    \textbf{Activation principles.} Given a feature $f$ to be activated:
    \begin{enumerate}
      \item \actmand: Activate all mandatory children of $f$.
      \item \actpar: If $g$ is $f$'s parent feature and $g$ is not a core feature, activate $g$.
      \item \actreq: If $f$ requires another feature $g$ (via a \textit{requires} relation) and $g$ is not a core feature, activate~$g$.
      \item \actgroup: If $f$ is an ``or'' or ``xor'' group,  activate one of $f$'s children.
      \item \actxor: If $f$ is a feature in an ``xor'' group,  deactivate all of $f$'s siblings.
      \item \actexcl: If $f$ excludes or is excluded by a feature $g$,  deactivate~$g$.
    \end{enumerate}
    
    \textbf{Deactivation principles.} Given a feature $f$ to be deactivated:
    \begin{enumerate}
      \item \dechild: If $f$ has a non-empty set of active children $G$ (including group members, optional, and mandatory children),  deactivate each child in~$G$.

      \item \dexor: If $f$ is a feature in an ``xor'' group,  activate one of $f$'s siblings or deactivate $f$'s parent unless it is core.
	  \item \deor: If $f$ is a feature in an ``or'' group, activate one of $f$'s siblings or deactivate $f$'s parent unless it is core.
      \item \deparent: If $f$ is a mandatory feature,  deactivate $f$'s parent.
      \item \dereq: If a feature $g$ requires $f$ (via  \textit{requires} relation),  deactivate~$g$.
    \end{enumerate}

  In general, a CPCO can consist of several \textit{variants} that arise from multiple ways of applying a particular principle (e.g., multiple siblings to choose from or deactivating the parent in \dexor).
	Therefore, the output of the generation procedure is a collection of variants.
		Executing the generated CPCO involves randomly picking and applying one variant.
	To produce all variants, the recursion is \textit{branched}:
	each way of addressing a principle is explored in a new call of the recursive procedure.
		When a branch ends (i.e., no further principle applications possible), the result is added to the collection of produced variants.

    Consider the generation of  the CPCO \textit{Act\_SMSTransfer} from Fig.~\ref{fig:ActSMSTransfer}.
    The initial solution contains only the activation node for \textit{SMSTransfer}.
    During successive, recursive calls, the \actmand\ and \actreq\ principles are applied, leading to partial and eventually full solutions that incorporate activation nodes for \textit{ReceivePhoto}, \textit{SendPhoto}, and \textit{CopyMedia}.

	Let us consider a CPCO with multiple variants.
    When generating \textit{De\_Screen3}, there are two possible ways of implementing principle \textsc{DeXor}: 
		we can activate either \textit{Screen1} or \textit{Screen2}.
		To ensure completeness, we fork the generation process and generate two separate rules, one of them leading to the activation of \textit{Screen1}, the other to the activation of \textit{Screen2}.
		These rules are equivalent to those shown in Fig.~\ref{fig:VBruleInstance}.

		If multiple groups and cross-tree constraints are involved, combinatorial effects quickly make this approach infeasible for large feature models.
		For example, for 2 alternative groups $f, g$ that each have 100 leaf children, the activation operator for $f$ comprises $100^2$ variants if $f$ requires $g$. 
    In the next section, we present an efficient generation algorithm that relies on the (de)activation principles, but addresses the problems highlighted here.

\section{Efficiently Generating CPCOs}
\label{section:generating_cpcos}

    %

  Generating CPCOs for realistic feature models becomes challenging quickly: 
  \begin{enumerate}
    \item Applying one of the (de)activation principles creates new feature decisions, requiring recursive application of the principles. 
          This can lead to potentially long sequences of feature decisions.
          Due to cyclic dependencies, a na\"ive generation approach might not terminate.
    \item Application principles like \actgroup{} or \deor{} offer alternative options for repair. 
          As the number of feature groups grows, the number of repair options multiplies.
  \end{enumerate}
  
  It may, therefore, not always be possible in practice to generate complete CPCOs that encode all possible variants of sustaining consistency in response to a given feature decision.
  Moreover, explicitly enumerating even a subset of variants may be inefficient. 
  In this section, we present an efficient technique for representing and generating a CPCO for a given feature decision. 
  We aim to generate `minimal' CPCOs capturing minimal configuration changes required to ensure consistency of the configuration.
  
	Our solution relies on the set of principles of feature activation and deactivation introduced in Sect.~\ref{section:cpcos}.
  In addition, three key ideas underlie our solution:
  \begin{enumerate}
    \item To efficiently compute the dependencies between \emph{all} feature decisions for a given feature model, avoiding the need to encode each possible path individually, we introduce the new concept of a \emph{feature-activation diagram}; 
    \item To minimise the size of the generated operators, we compactly encode CPCOs as VB rules (cf. Sect.~\ref{sec:example}, \cite{struber2018variability}); 
    \item To generate VB rules encoding a minimal CPCO for each feature decision, we efficiently analyse sub-diagrams of feature-activation diagrams.
  \end{enumerate}
  
  Figure~\figref{activity_diagram} gives an overview of the overall algorithm as an activity diagram. We will discuss the various steps in more detail below, starting with the notion of feature-activation diagrams and how they can be computed.
  
  \insertFigure[caption={Overview of CPCO generation algorithm.
                         This algorithm is invoked twice for every non-core, non-dead feature, with \texttt{activate} set to \texttt{true} and \texttt{false}, respectively.
												                         Grey activities are used to produce more efficient CPCO encodings by discarding unnecessary VB-rule instances.},
                onecolumn=T]%
               {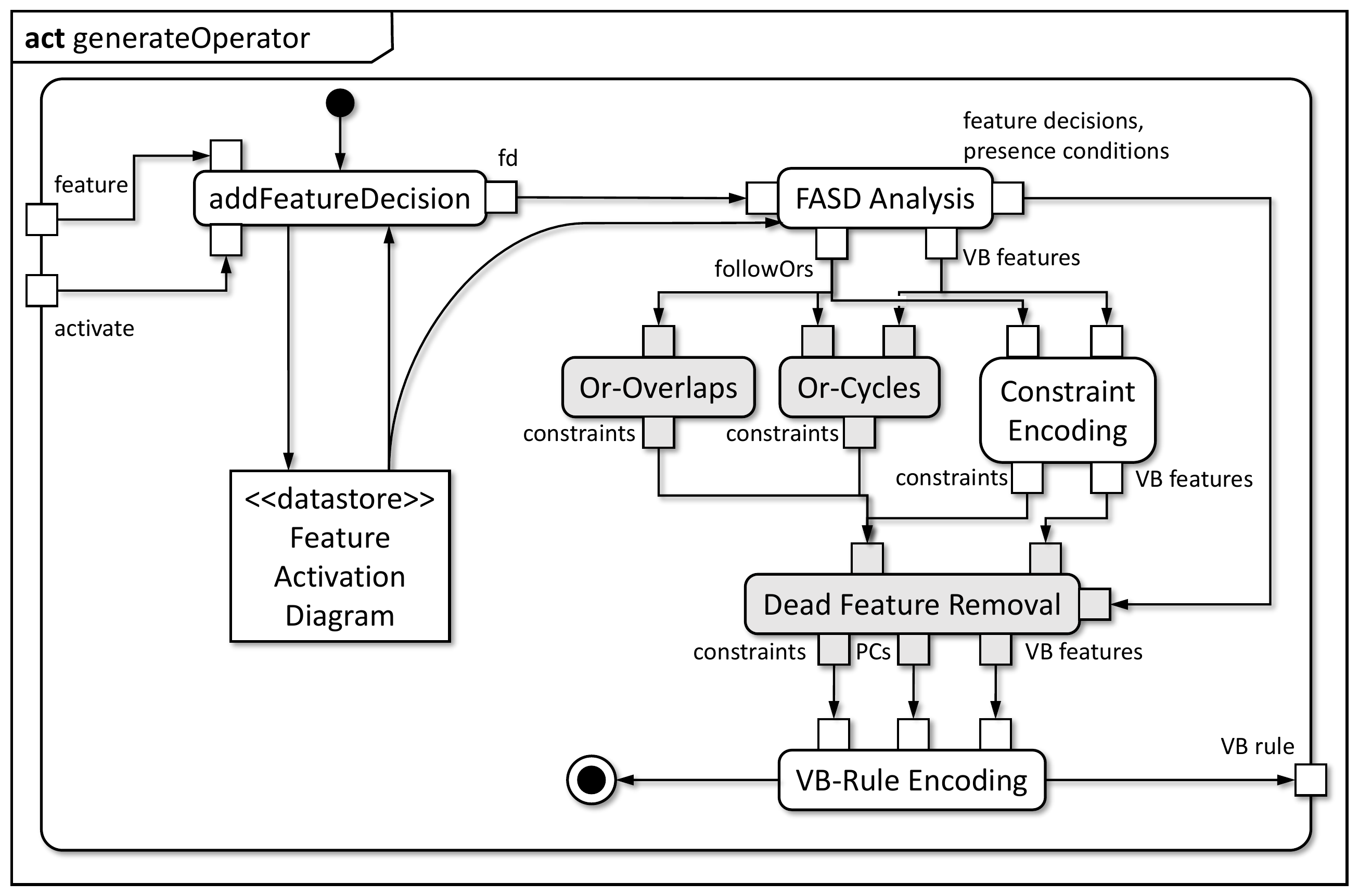}
  
	
  %
	%
    %

  \subsection{Feature-activation diagrams}
  
    A feature-activation diagram is a compact representation of all the implications of every feature decision for a given feature model.  
    These diagrams allow us to encode, in a graph, many often overlapping repair paths. The graph grows linearly with the number of real-optional features.
    Analysing feature-activation diagrams can efficiently produce CPCOs encoded as variability-based rules.
    
    
    \begin{definition}[Feature-activation diagram]\label{def:fad}
      A feature-activation diagram is a directed graph where the vertices are either:
      \begin{enumerate*}
        \item feature decisions, or
        \item or-nodes introducing alternative repairs.
      \end{enumerate*}
      The edges of a feature-activation diagram indicate direct consequences of a given feature decision. 
      Edges from feature decisions can lead to any kind of vertex.
      Edges from or-nodes can lead to feature decisions only.
    \end{definition}
    
    A complete feature-activation diagram for a given feature model contains exactly $2N$ feature decisions, where $N$ is the number of real-optional features in the feature model.
    Figure~\figref{fad-example} shows an example feature-activation diagram for a simple feature model.
    For example, the diagram shows that deactivating feature $F1$ requires the deactivation of features $F2$ and $F3$. Deactivating $F2$ requires either the deactivation of feature $F1$ or the activation of feature $F3$ and so on.
    
    \insertFigure%
      [caption={Example feature-activation diagram. 
                The bottom-left corner shows the feature model in FODA notation.
                The remainder of the figure shows the corresponding complete feature-activation diagram. 
                $+$/$-$ in feature-decision nodes indicates feature activation / deactivation, respectively.
                Arrows indicate the direct implications of a feature decision as a result of applying the (de)activation principles.
                Two example \emph{paths} are indicated in green (regular dashes) and orange (dash--dot)---see text for details.},
       onecolumn=T]%
      {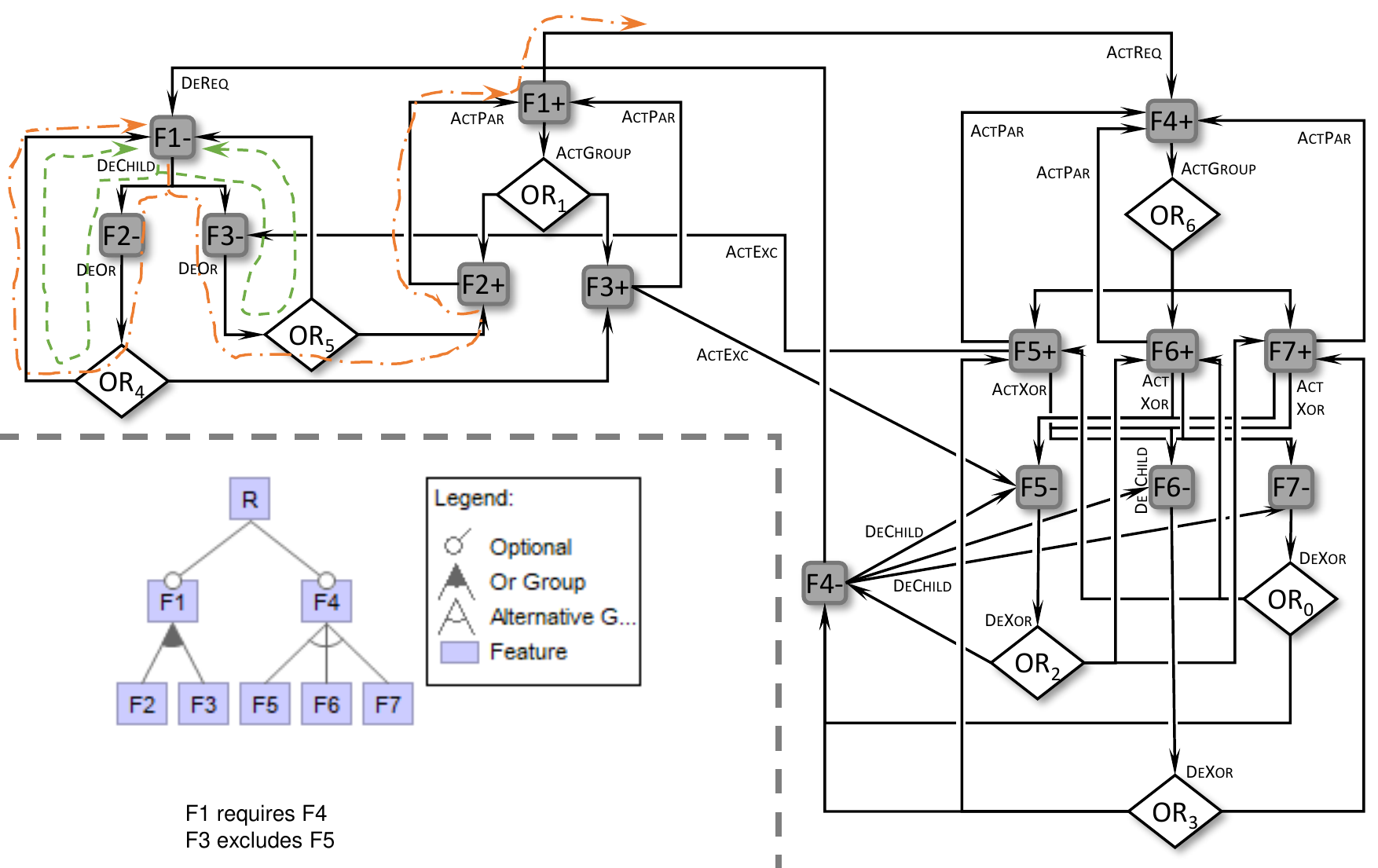}

    For a given feature decision, all feature decisions that can be reached on a \emph{path} describe one possible way of ensuring a consistent configuration as a result of the given feature decision.
    Multiple edges from a feature decision indicate that all vertices reachable in this way must be part of the path.\footnote{Note that this deviates from the common understanding of paths, which does not support forks. In the appendix, we define a notion of ``toggle graph'' that precisely captures our idea of paths.}
    Edges from an or-node indicate that only one of the consequences needs to be part of the path.
    As a result, or-nodes induce multiple valid paths through a feature-activation diagram.
    In Fig.~\figref{fad-example}, two example paths for $F1^-$ are shown with the green and orange dashed arrows, respectively.
    The green path indicates that one way of legally deactivating $F1$ is to also deactivate both $F2$ and $F3$ (and, as a transitive consequence, $F1$ again).
    The orange path also indicates this, but states that there is the option to attempt to make the deactivation of $F3$ legal by \emph{activating} both $F2$ and $F1$ (and further consequences from this, which we are not showing here explicitly).
    
    Clearly, this path contains contradictory decisions about the activation and deactivation of features. 
    We will \emph{not} remove such paths from the feature activation diagram, because this would require explicitly exploring every possible path.
    Instead, we will later (cf.~Sect.~\ref{section:generating_cpcos:analysing_fasds}) add constraints that ensure only consistent paths can be selected.
    
    Feature-activation diagrams can be efficiently computed in an incremental fashion using Algorithm~\ref{alg:fads}. 
    This corresponds to the activity labelled \texttt{addFeatureDecision} in Fig.~\figref{activity_diagram}.
    Starting from an arbitrary feature decision, we recursively apply the appropriate activation and deactivation principles from Sect.~\ref{section:cpcos}~(Line 7). 
    At each step, we add edges to represent all the newly computed consequences~(Lines~13--18).
    If a feature decision reached in this way is already part of the feature-activation diagram, the edge connects to that feature-decision node~(Lines~2--4).
    Otherwise, we create a new feature-decision node in the feature-activation diagram~(Lines~5--11).
    `Or' nodes are generated only for principles \actgroup{}, \dexor{}, or \deor{}.
    All consequences generated in this manner are added to the activation diagram~(Line~17).
  
    \begin{algorithm}[tbp]
      \caption{Generating feature-activation diagrams.}\label{alg:fads}
      \begin{algorithmic}[1]
        \scriptsize
        
        \Require{\textit{feature}: feature to be (de)activated.}
        \Require{\textit{activate}: Boolean indicating whether to activate or deactivate \textit{feature}.}
        \Require{\textit{diagram}: the global feature activation diagram.}
        \Ensure{Returns a FeatureDecision object that is contained in \textit{diagram} and represents the specified feature decision.}
        \Function{addFeatureDecision}{\textit{feature}, \textit{activate}}

          \State \textit{fdNode} $\gets$ \Call{find}{\textit{diagram}, <\textit{feature}, \textit{activate}>}
          \If{\textit{fdNode} found} \Return{\textit{fdNode}} \EndIf

          \State \textit{decisionNode} $\gets$ \textbf{new} FeatureDecision(\textit{feature}, \textit{activate})
          \State \textit{diagram} $\gets$ \textit{diagram} $+$ \textit{decisionNode}
          
          \LineComment \Call{applyPrinciples}{} computes the direct consequences of the given feature decision using the principles from Sect.~\ref{section:cpcos}.
          \State \textit{immediateConsequences} $\gets$ \Call{applyPrinciples}{\textit{decisionNode}}
          
          \ForEach{\textit{consequence} $\in$ \textit{immediateConsequences}}
            \State \Call{addConsequencesTo}{\textit{decisionNode}, \textit{consequence}}
          \EndFor
          \State \Return{\textit{decisionNode}}
        \EndFunction
        
        \Statex
        
        \Require{\textit{node}: a FeatureDecision contained in the global feature-activation diagram, possibly not yet linked to its consequences.}
        \Require{\textit{consequence}: set of or- and and-connected feature decisions.}
        \Ensure{\textit{node} is correctly linked to its consequences and all paths from it are included in the global feature-activation diagram.}
        \Procedure{addConsequencesTo}{\textit{node}, \textit{consequence}}          
          \ForEach{\textit{fd} $:$ \textsc{FeatureDecision} $\in$ \textit{consequence}}
            \State \textit{consequence} $\gets$
            \Statex $\qquad\qquad \left(\textit{consequence} \setminus \left\{\textit{fd}\right\}\right) \cup \{\Call{addFeatureDecision}{\textit{fd.feature}, \textit{fd.activate}}\}$
          \EndFor
          
          \State \textit{diagram} $\gets$ \textit{diagram} $+$ edges from \textit{consequence}
        \EndProcedure
      \end{algorithmic}
    \end{algorithm}

  \subsection{Analysing feature-activation sub-diagrams to generate CPCOs}
  \label{section:generating_cpcos:analysing_fasds}
  

    Once we have a feature-activation diagram, individual CPCOs can be generated by analysing the set of paths from a specific feature decision (the CPCO's ``root'' feature decision).
    Enumerating all paths explicitly can be prohibitively expensive.
    However, we can collect a compact encoding of all paths in a single depth-first search of the feature-activation diagram, starting at the root feature decision:
    %
    \begin{enumerate}
      \item 
            All CPCO variants for the same root decision are encoded in one VB rule. 
            The VB rule's feature model expression captures the different possible repairs; every valid configuration conforms to one possible repair.
            The structure of the feature-model expression is designed to avoid enumerating the repair options explicitly.
      \item 
            All information required for the construction of the VB rule encoding is collected in a single, linear-time depth-first traversal of the feature-activation diagram.
      \item 
            We discard unnecessary VB-rule instances by improving the VB-rule feature constraints, aiming to achieve minimality of the CPCO encoded by a VB rule.
    \end{enumerate}
    We discuss each of these steps in more detail below.
    
    \mypar{Encoding CPCOs as VB rules}
      Figure~\figref{detailed_vb_rule_example} shows an example of a VB rule we would generate from a feature-activation diagram.
      The feature-activation diagram is shown on the left of the figure and is a subset of the diagram in Fig.~\figref{fad-example} where, for illustration purposes, we ignore the cross-tree constraints from the original feature model.
      The VB rule implements the operators for the deactivation of feature F1.
      
      \insertFigure[caption={Example VB rule generated for deactivating feature F1 given the feature-activation diagram on the left.}]%
                   {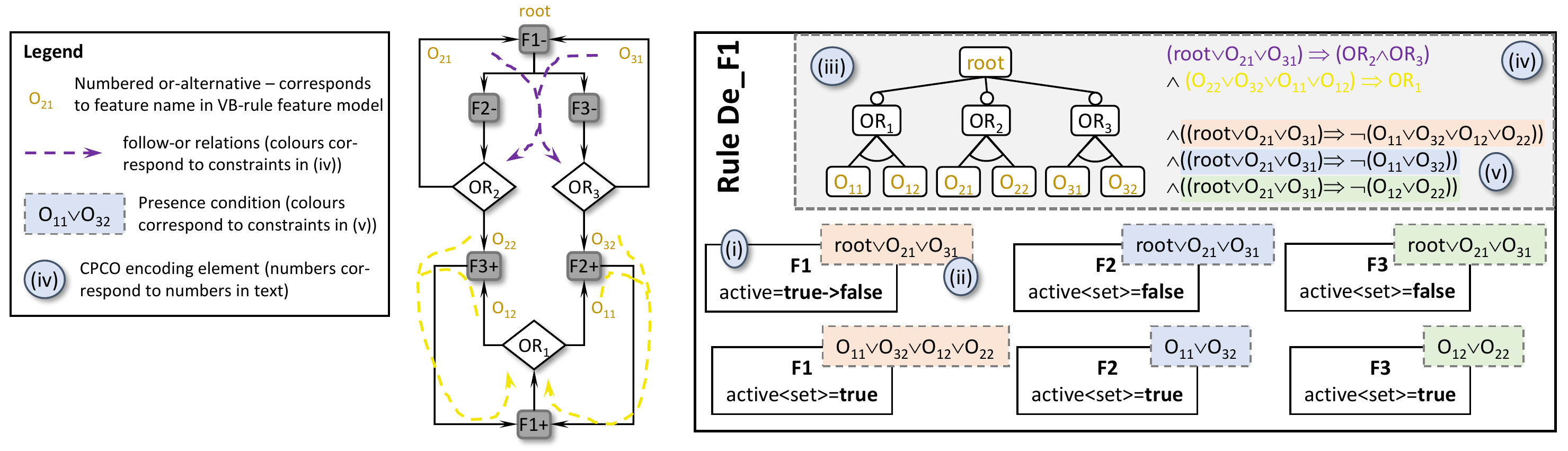}

      The key idea here is that we encode only the start of a path and the implications at each decision node in the feature-activation diagram rather than encoding every path individually.
      We encode CPCO variants using the following elements (the Roman numerals super-imposed over the rule in Fig.~\figref{detailed_vb_rule_example} correspond to the numbers in the following enumeration):
      \begin{enumerate*}[label={(\roman*)}]
        \item we include every feature decision from the activation sub-diagram (the root decision is marked up to check the original activation state of the feature and all other decisions are marked to simply change the state to the desired target state irrespective of the initial state); 
        \item we associate each feature decision with a presence condition, a disjunction of VB-rule features associated to paths that lead to the decision; 
        \item we add a top-level, \emph{optional} VB-rule xor-group feature for every or-node in the feature-activation sub-diagram with a child VB-rule feature for every edge leaving the or-node\footnote{Note that or-alternatives are labelled $O_{nm}$, where $n$ corresponds to the index identifiying the or-node and $m$ is an index uniquely identifying the alternative among the alternatives for or-node $n$.};
        \item \label{or-impls} we add cross-tree constraints to the VB rule that enforce an implication between each or-alternative and its \emph{next} or-node in the feature-activation diagram; and
        \item we add cross-tree constraints to the VB rule that disallow conflicting feature decisions (e.g., $F1^+$ and $F1^-$) to be selected at the same time (the constraints specify a mutual exclusion between the presence conditions of both feature decisions).
      \end{enumerate*}
      Point~\ref{or-impls} about or-implication constraints is particularly important: by only encoding the links between each or-alternative and the \emph{next} or-node, we avoid having to explicitly enumerate all paths.
      These are instead induced by the transitivity property of the implication operator.
      
      While the rule in Fig.~\figref{detailed_vb_rule_example} does indeed capture all possible variants for this case, it is unnecessarily complicated.
      For example, the exclusion constraints (v) mean none of the feature activations in the bottom row can ever be used, so they could be removed from the rule.
      In fact, the only possible configuration of the rule-specific feature model in this case is $\textit{root} \wedge O_{21} \wedge O_{31}$.
      We will return to this insight at the end of the section and describe how we generate more concise VB rules. 
      First, we describe how the information required can be efficiently extracted from a feature-activation diagram.
      Note that encoding is done after the information has been collected; Fig.~\figref{activity_diagram} shows this as an activity labelled \texttt{Constraint Encoding}.

    \mypar{Collecting CPCO information from feature-activation diagrams}
      We collect all information required for building the VB-rule encoding in a single sweep of the feature-activation sub-diagram.
      The key insight is that we can read off presence conditions by tracking the last or-nodes encountered and the branch taken out of those or-nodes, and we can read off the or-implications by tracking the next or-nodes we reach going forward through the feature-activation diagram.
      This can be done in a depth-first sweep: when we encounter a node we have previously visited, the information about any paths beyond that node is already available and we only need to add information about the new path through which the node can be reached (i.e., an additional presence condition). 
      To be able to take this additive approach and still propagate presence conditions through the graph, we track presence-conditions by proxy (\texttt{pc(x)} standing in for the actual final presence condition at node \texttt{x}, regardless of whether we have already collected all information required).
      These proxies can then be resolved after the depth-first sweep.
      
      This corresponds to the activity labelled \texttt{FASD Analysis} in Fig.~\figref{activity_diagram}. 
      Algorithm~\ref{alg:fasd-traversal} shows the core \texttt{visit} function invoked as part of the depth-first search.
      Starting from the root feature decision (Line~2), as we descend from a node to its follower nodes, we pass along the currently computed presence condition.
      As we return back up the graph, we return the last or-node seen; this is later used to construct the VB-rule or-implication constraints. 
      
      
      As discussed, or-implications and presence conditions are initially collected as proxies~(Lines~7 and 11, respectively) that require resolution after the traversal of the feature-activation sub-diagram is complete. 
      Resolving proxies requires a traversal of the underlying graph of proxies for each feature-decision node (not shown here).
      Cycles in the feature-activation diagram lead to cycles in the proxy chain, which are broken by removing any occurrence of a proxy in its own definition.
			
      \begin{algorithm}[tbp]
        \caption{Traversing feature-activation sub-diagrams.}\label{alg:fasd-traversal}
        \begin{algorithmic}[1]
          \scriptsize
          
          \LineComment{Collects the nodes in the feature-activation sub-diagram.}
          \State $\textit{subdiagramNodes} \in \mathcal{P}(\textit{ActivationDiagramNodes}) \gets \emptyset$
          \LineComment{Traverse the feature-activation diagram starting from the current \textit{root} feature decision.}
          \State $\textit{rootImplications} \gets$ \Call{visit}{\textit{root}, PC(`root')}
          
          \Statex
          
          \Require{\textit{node}: feature-activation-diagram node to visit.}
          \Require{\textit{pc}: presence condition collected so far (an object).}
          \Ensure{\textit{presenceConditions}: global map from feature decisions to list of presence-condition objects; initially empty.}
          \Ensure{\textit{featureDecisions}: global map from features to the feature-decisions encountered; initially empty.}
          \Ensure{\textit{globalFollowOrs}: global map from feature decisions to the or-nodes following them in the feature-activation diagram; initially empty.}
          \Ensure{Returns the set of or-nodes (or proxies) following \textit{node} in the feature-activation diagram.}
          \Function{visit}{\textit{node}, \textit{pc}}
            \If{\textit{node} is feature decision}
              \State $\textit{presenceConditions}[\textit{node}]$ += \textit{pc} 
              \Statex
              \If{$\textit{node} \in \textit{subdiagramNodes}$} 
                \LineComment Return a proxy object representing all or-nodes following the feature decision. 
                             This avoids having to explore from \textit{node} again by reusing information collected in other parts of the traversal. 
                \State \Return \Call{ProxyOrImplication}{\textit{node}} 
              \EndIf
              \State $\textit{subdiagramNodes} \gets \textit{subdiagramNodes} \cup \{\textit{node}\}$
              \Statex              
              \State $\textit{featureDecisions}[\textit{node}.\textit{feature}]$ += \textit{node}
              \Statex              
              \LineComment Return a proxy object representing all presence conditions collected for \textit{node}. 
                           Other parts of the traversal may reach \textit{node} and will add to the full presence condition represented by the proxy object.
              \State $\textit{newPC} \gets$ \Call{ProxyPC}{\textit{node}}
              \LineComment Step down. \textsc{flatMap} combines the sets returned into a single set.
              \State $\textit{followOrs} = \textit{node}.\textit{cons}.\textsc{flatMap}(n~|~\Call{visit}{n, \textit{newPC}})$
              \Statex
              \State $\textit{globalFollowOrs}[\textit{node}]$ += \textit{followOrs}
              \State \Return \textit{followOrs}

            \ElsIf{\textit{node} is or-node}
              \If{$\textit{node} \in \textit{subdiagramNodes}$}
                \State \Return \Call{featureFor}{\textit{node}} \Comment Find and return the VB-rule \Statex\Comment feature created for \textit{node}
              \EndIf
              \State $\textit{subdiagramNodes} \gets \textit{subdiagramNodes} \cup \{\textit{node}\}$
              \Statex
              \State $\textit{orFeature} \gets$ \Call{createFeatures}{\textit{node}} \Comment Create VB-rule feature for \textit{node} \Statex\Comment as an or-group with sub-features for each alternative
              \LineComment{Step down}
              \ForEach{$c \in \textit{or}.\textit{cons}$}
                \State $\textit{feature} \gets$ \Call{featureFor}{$c$}
                \State $\textit{followOnOrs} \gets$ \Call{visit}{$c$, \textsc{PC}(\textit{feature})}
                
                \State $\textit{globalFollowOrs}[c]$ += \textit{followOnOrs}
              \EndFor
              \Statex
              \State \Return \textit{orFeature}
            \EndIf
          \EndFunction
        \end{algorithmic}
      \end{algorithm}
      
      We, next, use the information thus gathered in the following way: 
      \begin{enumerate*}[label={(\roman*)}]
        \item for every entry in $globalFollowOrs$, we generate an implication $orAlternative \Rightarrow orFeature$ stating that or-node \texttt{orFeature} must be activated as a result of activating \texttt{orAlternative}, because there is a path from \texttt{orAlternative} to \texttt{orFeature}; and
        \item for every entry in $featureDecisions$ where both a positive and a negative feature decision have been included in the VB rule, we generate a mutual exclusion of the presence conditions of the two feature decisions.
      \end{enumerate*}

    \mypar{Discarding unnecessary VB-rule instances}
      The algorithm described will generate VB rules where every instance is a valid CPCO variant. 
      However, the resulting VB rules are unnecessarily large and the VB rule feature model allows an unnecessarily large number of rule instances, many of which are duplicates of other rule instances.
      We apply three improvements to the generated VB rule to address these problems.
      These improvements correspond to the activities labelled \texttt{Or-Overlaps} and \texttt{Or-Cycles} in Fig.~\figref{activity_diagram} as well as \texttt{Dead Feature Removal}.
      The first two of these improvements add additional constraints to the VB rule, while the third improvement makes the VB rule encoding more compact by removing parts of the rule that can never be instantiated.
      These additional steps can significantly improve the efficiency of the VB rules generated.
      For example, in WeaFQAs~\cite{Horcas2018WeaFQAs}, the initial VB rule generated for the deactivation of the \texttt{Security} feature allows more than 98,000 individual operators.
      After blocking self-activating cycles (see below), this is reduced to 320 rule instances.
      Removing dead features, removes 295 VB rule features, significantly reducing the rule size.
      We give only a high-level overview here, the full details are given in the supplementary material 
    (Appendix \ref{sec:appendix:cpco_gen}).
      
\begin{wrapfigure}{l}{.5\columnwidth}
      \includegraphics[width=.48\columnwidth]{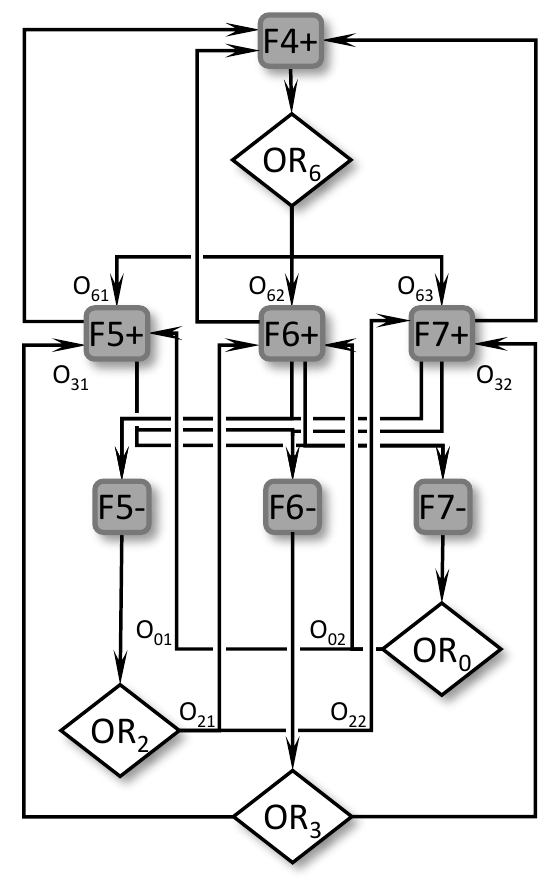}
      \caption{Example or-overlap.}
      \label{figure:or_overlap_example}
    \end{wrapfigure}

      \mysubpar{Constraints for or-overlaps} 
        Consider the excerpt of the feature-activation diagram from Fig.~\figref{fad-example} that is shown in Fig.~\figref{or_overlap_example}.
        Note how many or-nodes lead to the same feature decisions.
        For example, O$_{61}$, O$_{31}$, O$_{01}$ all lead to (F5$^+$). 
        This information is not captured in the VB rule we are currently generating and, as a result, the current VB rule feature model allows multiple configurations that lead to the same generated rule.
        For example, selecting O$_{61}$ and O$_{02}$ selects the same set of feature decisions as selecting O$_{62}$ and O$_{01}$.
        In addition to producing such redundant rule instances, we are also generating unnecessarily large repairs for a feature decision. 
        For example, if OR$_0$ is indirectly reached via O$_{62}$, then we have already made the decision to activate F6. 
        Activating F5 in addition to F6 does not improve the repair for the deactivation of F7 (which directly triggered OR$_0$), it just makes our operator larger---potentially a lot larger depending on the consequences of activating F5.
        We avoid such situations by adding explicit constraints that correlate decisions by different or-nodes.
        Continuing our example above, we would add a constraint $O_{62} \wedge OR_0 \implies O_{02}$ to say that we will always choose O$_{02}$ (and thus F6$^+$) if we activate OR$_0$ and have already activated O$_{62}$.
        Generating the additional data needed for constructing these constraints can be done as part of Algorithm~\ref{alg:fasd-traversal}.
        The key idea here is that we are already collecting data about the last or-alternative seen as part of collecting information for presence conditions.
        This information can be further analysed to identify or-overlaps.
        
        \mysubpar{Blocking of self-activating cycles} 
              Figure~\figref{or_overlap_example} also demonstrates another problem: self-activating cycles.
              In generating the VB-rule feature model, we have been able to avoid having to enumerate all repair paths by only encoding direct implications between or-alternatives and the directly following or-nodes and relying on the transitive nature of the logical implication to correctly reconstruct the paths on rule instantiation.
              However, this construction allows some superfluous rule instances to be constructed, too, namely, where there are cycles in the or-implications.
              For example, O$_{22}$ requires activation of OR$_6$ (because F$7^+$ leads to F4$^+$) and, in turn, O$_{62}$ requires activation of OR$_2$ (because F6$^+$ leads to F5$^-$).
              With the VB-rule feature model so far, activating or-alternatives in a cycle is always possible, even without a path from the root decision.
              This produces unnecessary repairs, making the operator unnecessarily complex.
              Because there are many cycles in a feature-activation diagram, and every cycle can be activated freely, we end up producing an unnecessarily large number of rule instances.
              To fix this, we add constraints to ensure cycles can only be activated if there is a path from the root decision into the cycle.

        \mysubpar{Removal of dead VB rule features} 
              We identify dead VB rule features and remove them from the VB rule.
              As a result, some feature decisions will have an empty presence condition, indicating they can never be part of any instance of the VB rule. 
              We remove these feature decisions from the VB rule.
			
			\subsection{Properties}
			\label{sec:algo_properties}


    We discuss three key properties of the generated CPCO suite.
		
\mypar{Soundness}	
Soundness is vitally important for us: Since our goal is to avoid the computation of fixes ``online'' during the optimization run, applying any generated CPCO to a valid configuration must not introduce constraint violations.
In appendix \ref{sec:appendix:soundness}, we provide a detailed and precise soundness argumentation.
Here, we give a summary of the main ideas.

During the first step (Algorithm 1), we recursively include fixes for all violations that might arise.
In consequence, each path of the feature activation diagram encapsulates a sound set of changes.
In the second step (Algorithm 2), we show that each instance of each generated CPCO represents a complete path from the feature-activation diagram. 
Finally, we show that the additional activities (gray part of \figref{activity_diagram}) do not threaten soundness, as they only remove potential CPCO instances, but do not add new ones.

\mypar{Completeness}	
			A noteworthy question is whether the generated CPCO suite is complete:
			starting from a given initial configuration, can an exhaustive search over the entire search space be performed by applying generated CPCOs?
			While we are optimistic that this is the case for our algorithm, a formal proof is outside the scope of this work.
			However, in our experimentation, we limit the number of generated instances per CPCO, which in some cases could lead to a loss of completeness: one CPCO instance might undo changes performed by another instance that would be required to reach a certain configuration.
			In Sect.~\ref{sec:evaluation}, we will see that our approach still improves over the current state of the art, which does not offer any completeness guarantees either.
			
\mypar{Performance}	
  We now discuss key performance aspects, referring to appendix \ref{section:appendix_complexity} for a more detailed argumentation.
	In the first step (Algorithm 1), feature-activation diagrams assume that the consequences of a feature decision are independent of the context in which this feature decision was made (i.e., the specific path through which we have reached a feature decision does not affect the relevant consequences).
	Therefore, a feature decision only occurs in a feature-activation diagram once and its consequences only need to be computed once.
  Generating a full feature-activation diagram, then, requires the equivalent of one complete traversal of the diagram---in effect, like a depth-first search through the graph.
  Thus, the complexity is bounded by $O\left(F + C + G\cdot A_g^2 + X\cdot A_x^2\right)$, where $F$ is the number of real-optional features, $C$ the number of cross-tree constraints, $G$ the number of group features, $A_g$ the average size of feature groups, $X$ the number of xor-groups, and $A_x$ the average size of xor-groups in the feature model.
  The computational complexity of constructing feature-activation diagrams is, thus, dominated by the average size of group features, but polynomial overall.
  
  In the second step (Algorithm 2),
  reading off the information required for a CPCO is, at worst, another full depth-first traversal of the feature-activation diagram.
  In addition, we have to resolve proxies, which may require an iteration over all features, and we generate two CPCOs for every real-optional feature.
  This leads to an overall complexity approximately cubic in the number of real-optional features.
	In the additional activities (gray part of \figref{activity_diagram}), the computationally most expensive step is the dead-feature removal, which relies on SAT solving.
  However, dead-feature removal does not significantly impact the efficiency of the generated CPCOs and we could remove it---trading space of CPCO representation against speed of CPCO generation.
\section{Tool support for CPCO-based automated optimal configuration}
\label{section:evolutionary_search}

To support optimal multi-objective product line configuration based on our generated CPCOs, we developed a tool called \emph{\ACAPULCO}.
The distinguishing feature of \ACAPULCO\ are its mutation and crossover operators, which are completely based on CPCOs, and guarantee solution validity throughout the search.
We use the genetic algorithm IBEA, based on an available implementation from the jMetal framework \cite{Durillo2011jmetal}. 
In \ACAPULCO's implementation, we reuse code fragments from \SATIBEA~\cite{satibea}, specifically, parts of its solution encoding, initial solution generation, and selection.
We replace \SATIBEA's most significant components---specifically, its mutation and crossover operators---with CPCO-based ones.

\mypar{Solution encoding} Our encoding consists of two variables: The main variable is
a bit vector, in which each bit represents the activation status of a particular feature.
This variable is based on the available encoding from IBEA.
Additionally, we maintain a \emph{history} of CPCO rules that were previously applied to produce the solution.
The history variable enables the use of CPCOs during crossover (explained below).

\mypar{CPCO generation} We generate CPCOs using the algorithm outlined in Sect.~\ref{section:generating_cpcos}.
As argued there, our generation algorithm is efficient.
Still,  there is a performance bottleneck further down the pipeline, as the generated CPCOs need to be \textit{instantiated} (via VB-rule configuration) to derive the encoded CPCO variants.
For this task, we rely on a SAT solver, which can rapidly produce individual configurations, but not \textit{all} configurations.
To keep the overall computational load tractable, we made two design choices:
we limited the number of generated variants per CPCO to 1, and used a time limit for CPCO generation, set to 10 minutes in our experiment.
Both design choices may affect the performance of the resulting CPCO suite.
Having more variants per CPCO leads to more alternatives to choose from in each iteration and, therefore, to a trade-off: in well-performing runs,  it may improve solution quality, while, across several runs, it may negatively affect robustness.
The time limit primarily affects completeness: for any features for which no (de)activation CPCO was generated before the time limit, that CPCO would be missing.
However, the (de)activation of these features could still be contained in other features’ CPCOs.


\mypar{Mutation and crossover} 
To mutate a given solution, we randomly pick one of its real-optional features and apply the corresponding activation or deactivation CPCO variant.
We map elements from the CPCO variant to bits from our encoding based on feature names (which we maintain as additional meta-information for the encoding).
For crossover, to derive two ``children'' from two given ``parents'', we copy each parent and apply the  CPCO variants from the history of the other parent.
We only apply those  CPCO variants not already included in the child's history.

\mypar{Further components}
To keep our comparative evaluation as fair as possible, we kept the other components close to SATIBEA's implementation.
For initial population generation, we incorporated \SATIBEA's and \MODAGAME's strategies of generating solutions by randomly applying one of multiple SAT solving strategies.
We reuse \SATIBEA's method for comparing and evaluating solutions based on the problem-specific objectives, formulated over the quality attributes of the problem instance, plus a heuristic ``helper'' objective (number of deactivated features).
Finally, we use \SATIBEA's \textit{binary tournament} selection operator.

\mypar{Quality assurance} 
We extensively tested our implementation on small test models as well as on all models from our evaluation, with up to 14K features.
In our test runs, our implementation behaved in line with the soundness guarantee: it never produced an invalid solution.

\section{Evaluation}
\label{sec:evaluation}
Our evaluation is based on ten standard benchmark feature models.
We studied the solution quality and execution time of automated configuration in our tool \ACAPULCO, compared to two state-of-the-art approaches and their associated tools.
All experiment results and artifacts to replicate the evaluation are available as an online appendix~\cite{appendix}.

\subsection{Experimental setup}
\label{sec:setup}

\mypar{Considered approaches}
\label{toolsForCompaison}
We compare against two state-of-the-art approaches for multi-objective optimization in SPLs and their associated tools: \MODAGAME~\cite{modagame} and \SATIBEA~\cite{satibea}. 
Both implement IBEA, which previously has been found to be the best-performing genetic algorithm for automated configuration~\cite{sayyad2013scalable,sayyad2013value,modagame,siegmund2017attributed,GuoICSE18}.
Both approaches rely on repair strategies, rather than avoiding invalid configurations.
To enable a fair comparison, we made the following adaptations:

\mysubpar{Quality attributes}
Both existing tools consider three fixed quality attributes.
\MODAGAME\ uses the floating-point attributes
\emph{usability},
 \emph{battery consumption}
and \emph{memory footprint}. 
\SATIBEA\ uses the attributes \emph{used before} (boolean),
 \emph{known defects} (float)
and \emph{cost} (float).
To compare all tools on a common set of attributes, we decided to use those of \MODAGAME, and replace the three of \SATIBEA; we modified \SATIBEA's implementation accordingly.
 We generated the attribute values for each feature in the same ranges as in MODAGAME's case studies based on a random uniform distribution, using \MODAGAME's random quality attribute generator.
 This type of distribution is a common practice in prior works for attribute values generation \cite{xiang2020going}. 

%



\mysubpar{Algorithm and parameter settings} \MODAGAME\ supports several optimization algorithms.
From analyzing the evaluation results reported in~\cite{modagame}, we identified  IBEA as the algorithm producing the best results for feature models with characteristics comparable to our evaluation models.
Hence, we used the IBEA implementation of \MODAGAME.
\SATIBEA\ and \ACAPULCO\ are also both based on IBEA.
We used the default parameters values for all tools (\eg\ mutation and cross-over probability for \MODAGAME\ and \SATIBEA), reported in the relevant papers~\cite{modagame,satibea}.

\mysubpar{Termination criteria} \MODAGAME\ stops after a given number of evolutions (\ie\ generations), while \SATIBEA\ uses a timeout, and \ACAPULCO\ supports both. 
To ensure a fair comparison, we use the number of evolutions in our experiments.
We extended \SATIBEA\ accordingly.



\mypar{Evaluation corpus}
We selected a set of ten feature models (Table~\ref{tab:FMs}), varying in size and complexity.
These feature models have been described and used in papers published within the software product line community, and they are often used for evaluation purpose~\cite{modagame,Siegmund2013,knuppel2017there}.
By design, our technique is geared towards feature models with basic  cross-tree constraints (``requires'' and ``excludes'').  Hence, we preprocessed the feature models to remove constraints representing more general boolean formulas. We later discuss the implications of this preprocessing. 
The quality attribute values were the same across the experiments for all tools. 

\begin{table}[t]
	\caption{Feature models corpus used for evaluation, with number of features, group features (``or'' and ``xor''), core features, cross-tree constraints (CTCs), number of configurations (size of the search space), number of CPCOs generated, and time to generate all CPCOs.}
	\label{tab:FMs}
	\centering
	\begin{threeparttable}
	\scriptsize
	\setlength{\tabcolsep}{0.5pt}
	\definecolor{Gray}{gray}{0.85}
	\newcolumntype{h}{>{\columncolor{Gray}}r}
	\begin{tabular}{lrrrrrhh}
		\toprule
		\rowcolor{white}
		Feature model
	    & \#Features
		& \#Groups
		& \#Core
		& \#CTCs
		& \#Configs
		& \#CPCOs
		& Time
		\\
		
		\midrule
		Wget~\cite{Siegmund2013}
		& 17
		& 1
		& 2
		& 0
		& 8,192
		& 30
		& 0.24 s
		\\
		
		Tank war~\cite{Siegmund2013} 
		& 37
		& 8
		& 7
		& 0
		& 580,608
		& 60
		& 0.41 s
		\\
	
		Mobile media~\cite{figueiredo2008evolving}
		& 43
		& 7
		& 10
		& 3
		& 2,128,896
		& 66
		& 0.35 s
		\\

		WeaFQAs~\cite{Horcas2018WeaFQAs}
		& 179
		& 36
		& 1
		& 7
		& 2.93e24
		& 356
		& 130.61 s
		\\
		
		
		Busy Box~\cite{knuppel2017there}
		& 854
		& 8
		& 20
		& 67
		& 2.1e201
		& 1,338
		& 1.48 s
		\\
		
		EMB ToolKit~\cite{knuppel2017there}
		& 1179
		& 70
		& 78
		& 1
		& 2.6e118
		& 426
		& 10 min
		\\
		
		CDL ea2468~\cite{knuppel2017there}
		& 1408
		& 12
		& 4
		& 1281
		& 3e136
		& 2,560
		& 20.23 s
		\\
		
		Linux Distribution~\cite{knuppel2017there}
		& 1580
		& 10
		& 6
		& 247
		& 2.85e419\tnote{*}~
		& 3,148
		& 43.08 s
		\\
		
	    Linux 2.6~\cite{knuppel2017there}
		& 6353
		& 137
		& 51
		& 3208
		& 3.90e1672\tnote{*}~
		& 3,844
		& 10 min
		\\
	    
	    Automotive 2.1~\cite{knuppel2017there}
		& 14009
		& 1135
		& 1394
		& 531
		& 4.7e1260
		& 1,600
		& 10 min
		\\
		
		\bottomrule
	\end{tabular}
\begin{tablenotes}
    \item[*] upper bound estimation.
\end{tablenotes}
\end{threeparttable}
	\vspace*{-0.4cm}
\end{table}




\mypar{Metrics}
We compare \ACAPULCO, \MODAGAME, and \SATIBEA\ on two quality criteria: \emph{solution quality} and \emph{performance}.
Our quality measurement relies on hyper-volume (HV)~\cite{Zitzler1999}, a standard metric for evaluating multi-objective approaches.
Due to its desirable theoretical properties, HV is widely accepted and used as a metrics for evaluating optimization approaches~\cite{DBLP:journals/corr/abs-2005-00515}, including the papers that introduced the two compared tools~\cite{modagame,satibea}.
HV measures the volume in the objective space covered by the members of a Pareto front \wrt\ a given reference point~\cite{Zitzler1999}. The reference point can be found by constructing a vector with the best objective function values running the algorithm for a high number of evolutions (\eg\ 10000). In our experiments, we rely on the jMetal framework~\cite{Durillo2011jmetal} to normalize and calculate HV. Higher values for HV are desirable, because a wider set of non-dominated solutions can be obtained.
Our measure of performance is execution time.
	
A further metric of interest is the percentage of invalid solutions in relation to the overall solution set.
Both \SATIBEA\ and \MODAGAME\ can produce invalid solutions.
However, while \SATIBEA\ yields invalid solutions as part of its solution set, \MODAGAME\ only provides the valid solutions it found as part of the solution set (for large feature models like \texttt{Linux} and \texttt{Automotive} \MODAGAME\ is unable to report any solutions).
For \SATIBEA, we report the ratio of produced valid solutions from the overall solution set.
For \ACAPULCO, we have a check ensuring that indeed no invalid solutions are reported (e.g., due to implementation bugs).

\mypar{Set-up}
For each tool and feature model presented in Table~\ref{tab:FMs}, we performed 30 runs, and calculated the means, medians and standard deviations for HV and execution time quality indicators.
We use a population of 100 solutions and a termination criterion of 50 generations (5000 evolutions), respectively, for each of our experiments.
The experiments were performed on a desktop computer with Intel Core i9-9900K, 3.6 GHz, 32 GB RAM, Windows 10, and Java 13. 

For hypothesis testing, we applied the Mann-Whitney U test~\cite{Arcuri2014}, commonly used with randomized algorithms in software engineering. 
This allows us to check that the differences between the three tools are statistically significant rather than being due to the inherent stochastic nature of the search process.
We apply the test to compare both the HV-values and the runtime of the three algorithms. The null hypothesis is that the values of \ACAPULCO\ are equal to the values of the other tools, while for the alternative hypothesis we used a standard library: the \texttt{scipy.stats.mannwhitneyu} package of SciPy.org. which supports the valuation of one-sided alternative hypothesis (``greater'', ``less''). Therefore, our alternative hypothesis is that the values of aCaPulCO exceed the values of the other tools in case of the HV metric, and that the values of aCaPulCO are lower than the other tools in case of the runtime metric.
	We report $p$-values, where a value below 0.05 means that the comparison is statistically significant at the 95\% confidence level.
We also assessed the effect size of our comparisons by using the $A_{12}$ score (calculated using the R package \texttt{effsize}), following Vargha and Delaney's original interpretation \cite{vargha2000critique}:
	$A_{12}\!\approx\!0.56 = \textit{small}$;
	$A_{12}\!\approx\!0.64 = \textit{medium}$; and
	$A_{12}\!\gtrsim\!0.71 = \textit{large}$. 

\subsection{Results}\label{sec:results}

  Table~\ref{tab:Evaluation} gives an overview of the results of our experimental evaluation. 
  The table shows the results on HV and time for the full run of 50 evolutions, providing the median values over 30 runs, the standard deviation, as well as $p$-values for the comparison between the tools. 
  Figures~\ref{fig:MobileMedia2FM}--\ref{fig:linuxFM} provide a more detailed analysis of the search process for three exemplary cases---all other cases are similar to at least one of the selected ones.
	Detailed data for the other cases are available in the online materials~\cite{appendix}. 
  Sub-figures~(a) contrast how the different tools converge to solutions over multiple evolutions, while sub-figures~(b) contrast the wall-clock execution time required by the tool to compute this number of evolutions.
  The diagrams show median values calculated over the 30 runs. 
  
  \begin{table*}[t]
	\caption{Comparison of hypervolumes (HV) and execution times (in seconds) of the three tools for each feature model. Results show median (MD) and standard deviation (SD) values over 30 runs. Right hand side shows the results of applying the Mann-Whitney U test comparing \ACAPULCO\ with \MODAGAME\ and \SATIBEA, for HV and time, respectively.}
	\label{tab:Evaluation}
	\centering
	\begin{threeparttable}
	\footnotesize
	\setlength{\tabcolsep}{1.5pt}
	\begin{tabular}{l|rrrr|rrrr|rrrrr||rr|rr}
		\toprule
		& 
		\multicolumn{4}{c|}{\centering \ACAPULCO}
		& \multicolumn{4}{c|}{\centering \MODAGAME}
		& \multicolumn{5}{c||}{\centering \SATIBEA}
		& \multicolumn{2}{c|}{\centering \ACAPULCO\ HV is greater}
		& \multicolumn{2}{c}{\centering \ACAPULCO\ time is faster}
		\\
		& \multicolumn{2}{c}{\centering HV}
		& \multicolumn{2}{c|}{\centering Time}
		& \multicolumn{2}{c}{\centering HV}
		& \multicolumn{2}{c|}{\centering Time}
		& \multicolumn{2}{c}{\centering HV}
		& \multicolumn{2}{c}{\centering Time}
		& \multicolumn{1}{c||}{\centering Invalid}
		
		& \multicolumn{1}{c}{\centering \MODAGAME}
		& \multicolumn{1}{c|}{\centering \SATIBEA}
		& \multicolumn{1}{c}{\centering \MODAGAME}
		& \multicolumn{1}{c}{\centering \SATIBEA}
		\\
		
		Feature model 
		& \multicolumn{1}{c}{\centering MD} & \multicolumn{1}{c}{\centering SD}
		& \multicolumn{1}{c}{\centering MD} & \multicolumn{1}{c|}{\centering SD}
		
		& \multicolumn{1}{c}{\centering MD} & \multicolumn{1}{c}{\centering SD}
		& \multicolumn{1}{c}{\centering MD} & \multicolumn{1}{c|}{\centering SD}
		
		& \multicolumn{1}{c}{\centering MD} & \multicolumn{1}{c}{\centering SD}
		& \multicolumn{1}{c}{\centering MD} & \multicolumn{1}{c}{\centering SD}
		& \multicolumn{1}{c||}{\centering Sols.}
		
		& \multicolumn{1}{c}{\centering $p$-value} 
		& \multicolumn{1}{c|}{\centering $p$-value} 
		
		& \multicolumn{1}{c}{\centering $p$-value} 
		& \multicolumn{1}{c}{\centering $p$-value} 
		\\
		
		\midrule

		Wget
		& \hlcell 0.49  & 7.43e-4
		& 0.23 & 0.05
		
		&  0.49 & 7.64e-4
		& \hlcell 0.22  & 0.04
		
		& 0.43 & 1.51e-2
		& 0.35  & 0.05
		& 1\%
		
		& 0.47
		& \hlcell 1.43e-11
		
		& 0.99
		& \hlcell 2.42e-10
		\\

		Tank war
		& \hlcell 0.55 & 1.63e-3
		& \hlcell 0.24  & 0.05

		& 0.55 & 4.96e-3
		& 0.25  & 0.04
		
		& 0.39 & 3.79e-2
		& 0.37 & 0.04
		& 5\%
		
		& \hlcell 6.15e-10
		& \hlcell 1.44e-11
		
		& \hlcell 8.31e-8
		& \hlcell 2.42e-10
		\\
	
		Mobile media
		& \hlcell 0.47 & 1.23e-3
		& \hlcell 0.24 & 0.05

		& 0.46 & 2.46e-3
		& 0.26 & 0.04
		
		& 0.35 & 2.11e-2
		& 0.34 & 0.04
	    & 1\%
	    
	    & \hlcell 4.59e-7
		& \hlcell 1.44e-11
		
		& \hlcell 1.28e-9
		& \hlcell 2.42e-10
		\\

		WeaFQAs
		& \hlcell 0.38 & 1.90e-3
		& \hlcell 0.29 & 0.05

		& 0.24 & 1.61e-2
		& 0.52 & 0.05
		
		& 0.24 & 2.18e-2 
		& 0.39 & 0.04
	    & 27\%
	    
	    & \hlcell 1.44e-11
        & \hlcell 1.44e-11
		
		& \hlcell 2.42e-10
        & \hlcell 2.42e-10
		\\
		
		
		
	
		
		
	    Busy Box
		& \hlcell 0.37 & 2.32e-3
		& \hlcell 0.36 & 0.07
		
		& 0.32 & 3.47e-3
		& 1.14 & 0.06
		
		& 0.29 & 1.23e-2
		& 0.53 & 0.05
	    & 15\%
	    
		& \hlcell 1.44e-11
		& \hlcell 1.44e-11
		
		& \hlcell 1.44e-11
		& \hlcell 2.42e-10
		\\
		
		EMB ToolKit
		& \hlcell 0.35 & 2.10e-3
		& \hlcell 0.61 & 0.08
		
		& 0.27 & 4.52e-3
		& 1.81 & 0.09
		
		& 0.29 & 1.07e-2
		& 0.80 & 0.08
	    & 26\%
	
		& \hlcell 1.44e-11
		& \hlcell 1.44e-11
		
		& \hlcell 1.44e-11
		& \hlcell 2.42e-10
		\\
		
		CDL ea2468
		& \hlcell 0.35 & 1.03e-3
		& 0.82 & 0.08
		
		& 0.17 & 5.91e-3
		& 4.16 & 0.07
		
		& 0.29 & 0.01
		& \hlcell 0.69 & 0.06
		& 23\%
	
		& \hlcell 1.44e-11
		& \hlcell 1.44e-11
		
		& \hlcell 1.44e-11
		& 0.99
		\\
		
		Linux Distrib.
		& \hlcell 0.34 & 2.33e-3
		& \hlcell 0.52 & 0.07
		
		& 0.31 & 2.26e-3
		& 2.44 & 0.08
		
		& 0.30 & 1.17e-2
		& 0.65 & 0.07
	    & 17\%
	
		& \hlcell 1.44e-11
		& \hlcell 1.44e-11
		
		& \hlcell 1.44e-11
		& \hlcell 2.42e-10
		\\
		
		Linux 2.6
		& \hlcell 0.33 & 3.73e-3
		& \hlcell 1.45 & 0.14
		
		& - & -
		& - & -
		
		& 0.29 & 9.50e-3
		& 1.88 & 0.12
		& 35\%
	
		& -
		& \hlcell 1.44e-11
		
		& -
		& \hlcell 2.20e-10
		\\
		
		Automotive 2.1
		& \hlcell 0.30 & 1.01e-2
		& \hlcell 19.49 & 0.89

        & - & -
		& - & -
		
		& 0.02 & 6.34e-2
		& 31.25 & 0.78
	    & 97\%
	
        & -
		& \hlcell 1.44e-11
		
        & -
		& \hlcell 1.44e-11
		\\
	    
		\bottomrule
	\end{tabular}
\begin{tablenotes}
	\item [] Runs: 30. Population: 100. Generations: 50 (5000 evolutions). \colorbox{gray!60}{Highlighted} the best results for hypervolume (highest value) and execution time (lowest value). If the values are equal, we highlight those with the lowest standard deviation. For $p$-values, we shaded cells below 0.05, i.e., the alternative hypothesis is true and the comparison is statistically significant at the 95\% confidence level.
\end{tablenotes}
\end{threeparttable}
 \vspace{-0.4cm}
\end{table*}
    
\begin{figure}
  \centering
  \begin{subfigure}[t]{\PLOTSUBFIG}
    \centering
    \begin{tikzpicture}[scale=\PLOTWIDTH]
        \begin{axis}[
        xlabel={Evolutions},
        ylabel={Hypervolume},
        xmin=200, xmax=5000,
        ymin=0.0, ymax=0.6,
        legend pos=south east
        ]
        
        \addplot [\MDEOPLOTCOLOR,\MDEOPLOTSTYLE,\PLOTGENERALSTYLE] table [x=NFE, y={HV Median}, col sep=comma]{plotsdata/Acapulco_mobile_media2_statsResults.dat};
        \addlegendentry{Acapulco}
        \addplot [\MODAGAMEPLOTCOLOR,\MODAGAMEPLOTSTYLE,\PLOTGENERALSTYLE] table [x=NFE, y={HV Median}, col sep=comma]{plotsdata/Modagame_mobile_media2_statsResults.dat};
        \addlegendentry{Modagame}
        \addplot [\SATIBEAPLOTCOLOR,\SATIBEAPLOTSTYLE,\PLOTGENERALSTYLE] table [x=NFE, y={HV Median}, col sep=comma]{plotsdata/Satibea_mobile_media2_statsResults.dat};
        \addlegendentry{Satibea}
        \end{axis}
    \end{tikzpicture}
    \vspace{-0.2cm}
    \caption{Hypervolume.}
    \label{sfig:MobileMedia2Vevols}
  \end{subfigure}
  \begin{subfigure}[t]{\PLOTSUBFIG}
    \centering
    \begin{tikzpicture}[scale=\PLOTWIDTH]
        \begin{axis}[
        xlabel={Evolutions},
        ylabel={Execution time (s)},
        xmin=200, xmax=5000,
        ymin=0, ymax=1,
        legend pos=north west
        ]
        
        \addplot [\MDEOPLOTCOLOR,\MDEOPLOTSTYLE,\PLOTGENERALSTYLE] table [x=NFE, y={Time Median (s)}, col sep=comma]{plotsdata/Acapulco_mobile_media2_statsResults.dat};
        \addlegendentry{Acapulco}
        \addplot [\MODAGAMEPLOTCOLOR,\MODAGAMEPLOTSTYLE,\PLOTGENERALSTYLE] table [x=NFE, y={Time Median (s)}, col sep=comma]{plotsdata/Modagame_mobile_media2_statsResults.dat};
        \addlegendentry{Modagame}
         \addplot [\SATIBEAPLOTCOLOR,\SATIBEAPLOTSTYLE,\PLOTGENERALSTYLE] table [x=NFE, y={Time Median (s)}, col sep=comma]{plotsdata/Satibea_mobile_media2_statsResults.dat};
        \addlegendentry{Satibea}
        \end{axis}
    \end{tikzpicture}
    \vspace{-0.2cm}
    \caption{Execution time.}
    \label{sfig:MobileMedia2ExecTime}
  \end{subfigure}
   %
\vspace{-0.1cm}
  \caption{Results for Mobile Media feature model.}
\vspace{0.3cm}
  \label{fig:MobileMedia2FM}
  \begin{subfigure}[t]{\PLOTSUBFIG}
    \centering
    \begin{tikzpicture}[scale=\PLOTWIDTH]
        \begin{axis}[
        xlabel={Evolutions},
        ylabel={Hypervolume},
        xmin=200, xmax=5000,
        ymin=0, ymax=0.6,
        legend pos=north east
        ]
 
        \addplot [\MDEOPLOTCOLOR,\MDEOPLOTSTYLE,\PLOTGENERALSTYLE] table [x=NFE, y={HV Median}, col sep=comma]{plotsdata/Acapulco_weafqas_statsResults.dat};
        \addlegendentry{Acapulco}
        \addplot [\MODAGAMEPLOTCOLOR,\MODAGAMEPLOTSTYLE,\PLOTGENERALSTYLE] table [x=NFE, y index={5}, col sep=comma]{plotsdata/Modagame_weafqas_statsResults.dat};
        \addlegendentry{Modagame}
         \addplot [\SATIBEAPLOTCOLOR,\SATIBEAPLOTSTYLE,\PLOTGENERALSTYLE] table [x=NFE, y index={5}, col sep=comma]{plotsdata/Satibea_weafqas_statsResults.dat};
        \addlegendentry{Satibea}
        \end{axis}
    \end{tikzpicture}
        \vspace{-0.2cm}
    \caption{Hypervolume.}
    \label{sfig:WeaFQAsHVevols}
  \end{subfigure}
  \begin{subfigure}[t]{\PLOTSUBFIG}
    \centering
    \begin{tikzpicture}[scale=\PLOTWIDTH]
        \begin{axis}[
        xlabel={Evolutions},
        ylabel={Execution time (s)},
        xmin=200, xmax=5000,
        ymin=0, ymax=1,
        legend pos=north west
        ]
        
        \addplot [\MDEOPLOTCOLOR,\MDEOPLOTSTYLE,\PLOTGENERALSTYLE] table [x=NFE, y={Time Median (s)}, col sep=comma]{plotsdata/Acapulco_weafqas_statsResults.dat};
        \addlegendentry{Acapulco}
        \addplot [\MODAGAMEPLOTCOLOR,\MODAGAMEPLOTSTYLE,\PLOTGENERALSTYLE] table [x=NFE, y={Time Median (s)}, col sep=comma]{plotsdata/Modagame_weafqas_statsResults.dat};
        \addlegendentry{Modagame}
        \addplot [\SATIBEAPLOTCOLOR,\SATIBEAPLOTSTYLE,\PLOTGENERALSTYLE] table [x=NFE, y={Time Median (s)}, col sep=comma]{plotsdata/Satibea_weafqas_statsResults.dat};
        \addlegendentry{Satibea}
        \end{axis}
    \end{tikzpicture}
        \vspace{-0.2cm}
    \caption{Execution time.}
    \label{sfig:WeaFQAsExecTime}
  \end{subfigure}
   %
   \vspace{-0.1cm}
  \caption{Results for WeaFQAs feature model.}\vspace{0.3cm}

  \label{fig:WeaFQAsFM}
  \begin{subfigure}[t]{\PLOTSUBFIG}
    \centering
    \begin{tikzpicture}[scale=\PLOTWIDTH]
        \begin{axis}[
        xlabel={Evolutions},
        ylabel={Hypervolume},
        xmin=200, xmax=5000,
        ymin=0, ymax=0.6,
        legend pos=north west
        ]
        
        \addplot [\MDEOPLOTCOLOR,\MDEOPLOTSTYLE,\PLOTGENERALSTYLE] table [x=NFE, y={HV Median}, col sep=comma]{plotsdata/Acapulco_linux-2.6.33.3_statsResults.dat};
        \addlegendentry{Acapulco}
         \addplot [\SATIBEAPLOTCOLOR,\SATIBEAPLOTSTYLE,\PLOTGENERALSTYLE] table [x=NFE, y={HV Median}, col sep=comma]{plotsdata/Satibea_linux-2.6.33.3_statsResults.dat};
        \addlegendentry{Satibea}
        \end{axis}
    \end{tikzpicture}
        \vspace{-0.2cm}
    \caption{Hypervolume.}
    \label{sfig:linuxHVevols}
  \end{subfigure}
  \begin{subfigure}[t]{\PLOTSUBFIG}
    \centering
    \begin{tikzpicture}[scale=\PLOTWIDTH]
        \begin{axis}[
        xlabel={Evolutions},
        ylabel={Execution time (s)},
        xmin=200, xmax=5000,
        ymin=0, ymax=3,
        legend pos=north west
        ]
        
        \addplot [\MDEOPLOTCOLOR,\MDEOPLOTSTYLE,\PLOTGENERALSTYLE] table [x=NFE, y={Time Median (s)}, col sep=comma]{plotsdata/Acapulco_linux-2.6.33.3_statsResults.dat};
        \addlegendentry{Acapulco}
        \addplot [\SATIBEAPLOTCOLOR,\SATIBEAPLOTSTYLE,\PLOTGENERALSTYLE] table [x=NFE, y={Time Median (s)}, col sep=comma]{plotsdata/Satibea_linux-2.6.33.3_statsResults.dat};
        \addlegendentry{Satibea}
        \end{axis}
    \end{tikzpicture}
        \vspace{-0.2cm}
    \caption{Execution time.}
    \label{sfig:linuxExecTime}
  \end{subfigure}
   %
   \vspace{-0.1cm}
  \caption{Results for Linux 2.6 feature model.}
  \label{fig:linuxFM}
\end{figure}




  
 \mypar{Solution quality}
       As shown by the HV data in Table~\ref{tab:Evaluation}, \ACAPULCO\ generally finds better solutions than \MODAGAME\ and \SATIBEA. In the three smallest cases, the median quality of the solutions found is similar to those of \MODAGAME{}.
			In the larger cases, it appears that \ACAPULCO\ is able to cover a larger part of the objective space than the other tools. The difference in HV becomes greater as the search space of the feature model grows in size, as occurs for \texttt{WeaFQAs} (Fig.~\ref{fig:WeaFQAsFM}): 0.38 covered by \ACAPULCO\ against 0.24 covered by \MODAGAME\ and by \SATIBEA\ (45\% of difference); and for \texttt{Linux} (Fig.~\ref{fig:linuxFM}): 0.33 covered by \ACAPULCO\ against 0.29 covered by \SATIBEA\ (13\% of difference). \MODAGAME\ is unable to find and report any solution for the \texttt{Linux} feature model.
			In the case of \textit{Automotive 2.1}, \SATIBEA\ only reports 3\% valid solutions and a substantially lower HV than \ACAPULCO.

			All three tools show small standard deviations between solutions, indicating a high robustness~\cite{dreschler2007bremen}.
			That is, the variance between the solutions found in different runs is small---an important criterion for practical use where it is not feasible to execute many runs and select the best results.    

			The observed quality differences are statistically significant, with the exception of the smallest  case \texttt{Wget}.
The differences between the tools are particularly pronounced in terms of effect sizes.
Except for case \texttt{Wget}, every comparison between \ACAPULCO\ and one of the compared tools exhibits a large effect size (0.86 $\leq$ $A_{12}$ $\leq$ 1.0, see Appendix~\ref{sec:appendix:effsize}).

To ensure that the observed difference does not come from a conveniently chosen termination criterion, we performed additional experiments in which all tools were executed with 20,000 instead of 5,000 evolutions (see Appendix~\ref{sec:appendix:add_exp}).
Yet, in these new experiments, the solutions found by \MODAGAME\ and \SATIBEA\ are still outperformed by the solutions found by \ACAPULCO.
			We conclude that \ACAPULCO\ produced improved solutions in terms of quality, even more so for larger feature models. 
    
    %
    
  \mypar{Convergence speed}
    Figures~\ref{fig:MobileMedia2FM}--\ref{fig:linuxFM} allow us to compare the convergence behavior of the three tools.
		We see that \ACAPULCO\ generally converges faster than \SATIBEA.
		The initial advantage can be explained because even at the beginning, all solutions reported by \ACAPULCO\ are valid, giving a greater HV than \SATIBEA's HV with partially invalid solutions.
		The same applies for \MODAGAME, which behaves almost identically to \ACAPULCO\ in the smallest case, \textit{MobileMedia}.
		On the two larger cases, the drawbacks of \MODAGAME{}'s strategy become more manifest, as it stagnates close to the initial HV for the case of WeaFQAs, and cannot find any solutions for Linux 2.6.
    %
 
   Our tool \ACAPULCO\ never produces an invalid configuration.
  As a result, it can start exploring the search space rather than spend time repairing the candidate solutions in its population, as \MODAGAME\ and \SATIBEA\ have to. Similarly, \ACAPULCO\ has good potential for moving out of local optima because its mutation and crossover operators contain self-contained sets of changes that inherently lead to other valid solutions.
   		Overall, we observe improved convergence for \ACAPULCO\ especially for large feature models.

    
  \mypar{Execution time}
	\ACAPULCO\ statistically significantly yields the best execution times of all three tools for almost all cases.
	The only exceptions are \texttt{Wget} and \texttt{CDL ea2468}, where \MODAGAME\ and \SATIBEA, respectively, are faster.
	The difference in time is  significant in medium and large search spaces as for \texttt{WeaFQAs} (Fig.~\ref{fig:WeaFQAsFM}): 0.29 seconds spent by \ACAPULCO\ against 0.52 seconds spent by \MODAGAME\ (57\% of difference), and 0.39 seconds spent by \SATIBEA\ (29\% of difference).
	\MODAGAME, in particular, has been highly optimised to be executable in a resource-limited mobile environment~\cite{modagame}, and this is clearly visible in the execution-time results for the smaller feature models like \texttt{Wget}, \texttt{Tank War}, and \texttt{Mobile Media}. However for larger feature models, \MODAGAME\ spends most of its execution time fixing invalid solutions. Its repair operator randomly toggles features until a valid configuration is found~\cite{modagame}. Thus, for large-scale feature models like \texttt{Linux} and \texttt{Automotive} the operator is not efficient or even unable to find a valid configuration.
The observed effects are strong:   
in 16 out of 18 comparisons, \ACAPULCO\ outperforms the compared tools with a large effect size (0.89 $\leq$ $A_{12}$ $\leq$ 1.0; see Appendix~\ref{sec:appendix:effsize}).

\subsection{Discussion}
\label{sec:discussion}

\mypar{Online vs. offline} A key feature of our approach is that the operator generation is performed ``offline'', before the actual search, compared to the existing approaches that compute repair steps during the search.
This saves redundant computation effort both during the search and across multiple search runs.
The latter is beneficial especially if the search is to be run repeatedly, for example, in a dynamic reconfiguration context in which the quality attribute values are monitored and change constantly, making it necessary to adapt the configuration to the changed values.

In a static application context where the search is executed only once, the offline step can constitute a substantial part of the total execution time.
The \textit{total} execution time of \ACAPULCO\  can then be worse than that of the compared tools.
For example, when interpreting our results under the assumption of such a ``one-shot'' optimization context, 
the total execution of \SATIBEA\ (31 seconds max.) is generally shorter than \ACAPULCO's offline step alone, which in three cases took the specified maximum time of 10 minutes (see the generation times in Table~\ref{tab:FMs}).
However, our approach still produces substantially better solutions.
In the largest considered case \textit{Automotive 2.1}, \ACAPULCO\ clearly outperforms \SATIBEA\ with a HV of 0.30 instead of 0.02, whereas \MODAGAME\ is not even able to produce a solution.

%

\mypar{Expressiveness} We assume a basic feature model dialect with all FODA~\cite{foda} concepts, namely: mandatory, optional, ``or'' and ``xor'' group features, and basic  cross-tree constraints (``requires'' and ``excludes''). 
In practice, the usage frequency and types of cross-tree constraints vary significantly between projects. For instance, in Berger et al.'s industry study~\cite{berger2013survey}, 80\% of the surveyed participants confirmed the existence of constraints in their projects, and 45\% state that on average only a minority (less than 25\%) of features is affected by constraints.
Constraints also may aim at different use cases (\emph{e.g.,} enforcing correctness \emph{vs.} improving user experience during configuration~\cite{nadi2015configuration}), which might require different levels of expressiveness.


The most significant constraint type we do not address explicitly are complex constraints based on arbitrary propositional formulas.
Such constrains are important, for example, in the automotive and embedded systems domains~\cite{knuppel2017there}.
This limitation does not apply to our compared tools SATIBEA \cite{satibea} and MODAGAME \cite{modagame}, whose repair operators are geared to support arbitrary propositional constraints.

Knüppel et al.~\cite{knuppel2017there} provide an algorithm for transforming complex constraints into additional features and basic constraints.
While this algorithm could render our method applicable to the relevant feature models, it significantly increases the number of features and constraints, creating a scalability challenge.
In additional exploratory experiments, we applied our CPCO generation algorithm to versions of our evaluation models created by that algorithm.
The CPCO generation scaled up to most considered cases, including the largest considered one \textit{Automotive 2.1}, but not to \textit{Linux 2.6}.
To support complex constraints in such cases, we propose a hybrid search strategy: use CPCOs for all basic constraints and SATIBEA's \textit{fix} operator \cite{satibea} to repair violations of remaining constraints.
Such a technique might yield a ``sweet spot'' by relying much less on arbitrary repair and its associated drawbacks than the compared tools do, while still using repair for those constraints that cannot be addressed with CPCOs yet.
This is especially promising for feature models that predominantly include basic constraints.
For example, Knüppel et al.\ report over 80\% of all constraints in the automotive domain to be basic~\cite{knuppel2017there}.


\mypar{Feature attributes}
We do not make any assumptions about feature attributes.
In our evaluation, 
we focused on the attribute model supported by the tools we compare against, which does not address feature interactions.
Studying the impact of CPCOs on problems with interacting features is an important avenue for future work, since feature interactions lead to different fitness landscapes than orthogonal ones~\cite{siegmund2017attributed}.
Since CPCOs lead to improved results in cases with basic attributes, an NP-complete problem~\cite{literatureReview}, we hypothesize that we will see improved results for problems with feature interactions as well.
Further extensions not addressed by our approach include clonable features~\cite{Czarnecki2005}, numerical features~\cite{Munoz2019}, as well as constraints over quality attributes \cite{garcia2016automated}.
We aim to consider these extensions in the future.

\subsection{Threats to validity}

\mypar{Internal validity}
A threat to internal validity is our choice of configuration parameters for the different tools.
In the experiments, for the considered tools, we used the default parameter values as documented in the associated papers~\cite{modagame,satibea}.
We believe that this setup ensures a fair comparison, since all approaches are treated in the same way.
While a detailed evaluation of the optimization tool configuration is out of the scope of this paper, we plan, as our future work, to study how the tools' configuration parameters  affect the search, and to perform a sensitivity analysis to study the robustness of the tools wrt. configuration parameters.

\mypar{External validity}
Threats to external validity concern the generalization of the results to other cases and tools.
First, the generalization to other cases might be limited by the expressiveness limitations identified and discussed in Sect.~\ref{sec:discussion}.
Second, like  various state-of-the-art works \cite{sayyad2013scalable,sayyad2013value,xiang2020going}, we rely on  randomly generated attribute values based on a uniform distribution, leaving a study of the impact of different distributions to future work.
The closest work in this direction is Siegmund et al.'s~\cite{siegmund2017attributed} performance comparison of a single method on different problem instances (attribute values sampled from a normal distribution \emph{vs.} a randomized distribution derived from empirical data).
Our scenario is different as we compare \textit{different} methods on the \textit{same} problem instance, so that the impact of the chosen attribute-value distribution is controlled for. 
Third, comparing our technique to a wider selection of tools would lead to more comprehensive results.
Still, our considered tools are widely used; specifically \SATIBEA\ has inspired follow-up studies fine-tuning aspects of its parametrization \cite{GuoICSE18,xiang2018configuring,xiang2020going}.
    
\mypar{Construct validity}
Implementation details can largely affect the performance of a tool. 
Since \MODAGAME\ is focused on reducing execution time to be executed in mobile devices~\cite{modagame}, its implementation relies on performance-optimized data structures from the High Performance Primitive Collections (HPPC) instead of the Java built collection library.
To address this potential noise variable, we intend to experiment with HPPC in the future as well.   

\mypar{Conclusion validity}
Conclusion validity relates to the reliability and robustness of our results. 
We address conclusion validity by executing 30 independent runs of each experiment and applying standard statistical analysis techniques (\eg\ Mann-Whitney U test).
Moreover, all the code, artifacts, and data used in these experiments are available for replication and further analysis~\cite{appendix}.

\section{Related Work}
\label{section:related_work}

 \mypar{Manual configuration}
Configuration of SPLs is a process that has been extensively studied. Some approaches on feature model configurations are based on support for the manual selection of features based on automated recommendations. These suggestions can be based on heuristics exploiting structural characteristics of the feature model (\eg\ the most constrained variables first)~\cite{recommendation}, information from existing configurations~\cite{recommendation,visualRecommendations,frogs}, or on how features impact non-functional properties~\cite{visualRecommendations}. Despite that the initial objective is to speed up the manual selection process, these recommendation systems can often be used to automatically complete the configuration process.

	Repair approaches, such as range-fixes~\cite{rangeFixes} and FaMa~\cite{fama}, both take an invalid configuration as input, and generate a list of possible fixes for the configuration.
	The user selects the preferred fix.
	Both approaches assume a given faulty configuration, and compute repair actions for it.
	In contrast, our approach generates an operator suite that preserves validity when applied to \textit{any} given valid configuration.

To support users in making correct choices during manual configuration, Krieter et al.~\cite{krieter2018propagating} present a technique for propagating the consequences of feature selection and deselection, based on modal implication graphs (MIGs).
		Similar to us, this approach tries to guide the configuration process to avoid invalid configurations, rather than fixing them.
		An important difference is that this technique does not aim to preserve validity.
		A technical reason why this is challenging is that MIGs represent group constraints in a coarse-grained way:
		MIGs have an edge type for modeling that two features are dependent ``under certain conditions'', which includes group constraints.
		Consequently, propagation may lead to violations of group constraints that have to be fixed manually by the user, leading to a semi-automated configuration process.
		In our approach,  we explicitly model the different options for dealing with group constraints using OR nodes.
		This allows full automation addressing all constraints while avoiding repair actions.

 \mypar{Automated optimal configuration} 
Ochoa et al.~\cite{literatureReview} present a systematic literature review on automated configuration.
The seminal approaches in the field relied on constraint-satisfaction problem (CSP) solvers~\cite{benavides2005automated,splconqueror} and custom heuristics~\cite{splconfig}.
 Examples of the now predominant search-based paradigm are \MODAGAME~\cite{modagame}, \SATIBEA~\cite{satibea}, and ClaferMoo~\cite{clafermoo}. 
To deal with validity constraints, these works rely on repair strategies being computed during the search.
\MODAGAME\ includes a custom  ``fix'' operator, whereas \SATIBEA\ uses a SAT solver within a ``smart'' mutation operator whose purpose is to remove violations introduced in previous mutations.
Xiang et al. vary \SATIBEA\ by studying the impact of different SAT solving techniques and configurations during repair \cite{xiang2018configuring,xiang2020going}.
We discuss the positioning of our work in this line of research in Sect.~\ref{section:introduction}.

	Preserving valid configurations is a desired property in automated optimal configuration.
	Guo and Shi~\cite{GuoICSE18} perform an experimental evaluation of different search strategies that differ in their handling of invalid solutions.
	For their experiments, they developed several variants of the \SATIBEA~\cite{satibea} tool.
	While they did not consider operators that ensure validity of candidate solutions as we do, they find favorable results for strategies that preserve valid solutions.

We assume that the quality attribute values are available. Inferring these values (that is, building a performance model) is a research direction by itself~\cite{Siegmund2013}, and an accurate performance model can help to significantly improve the performance of a given optimisation framework~\cite{nair2018faster}.
As an alternative to building a performance model, it has been proposed~\cite{oh2017finding,Batory2021} to use random sampling to directly infer configurations, and optimise them by identifying features contributing to improved performance.
This solution offers performance benefits when the attribute values are unknown.
The authors also report improved accuracy when finding optimal solutions during single-objective optimization.
In the present work, we consider multi-objective optimisation, and a situation where the attribute values are known.

 \mypar{Further automated analyses}
	Beyond the optimisation context, a related problem is to find \textit{arbitrary} valid configurations (as opposed to optimal ones)---a non-trivial issue for large feature models.
	The standard approach for finding an arbitrary configuration is to translate the  feature model into a propositional formula and feed it to a SAT solver.
	SAT solvers scale up to millions of variables~\cite{basicfm}.
	As an alternative when \textit{all} valid configurations need to be enumerated, BDDs are known to be particularly efficient \cite{benavides2005automated}, although they only support cases with thousands of features, which excludes the \textit{linux} and \textit{automotive} cases from our evaluation. SMT solvers have been proven useful for a case with more than half a million features \cite{lienhardt2020lazy}.
While our approach has similarities to the internal workings of a solver,
	SAT, BDD and, for that matter, CSP and SMT solving are complementary to our approach, since
	they produce concrete valid configurations, instead of providing bundles of changes that guarantee validity during reconfiguration. 
  We reuse standard SAT solvers for instantiating the VB rules we generate for CPCOs.

A related concept to our operators are \emph{atomic sets}~\cite{benavides}, which bundle several features that are always activated together.
The main use case of atomic sets is to make automated analysis of SPLs more efficient by treating each atomic set as a single, abstracted feature.
However, in the context of configuration, toggling entire atomic sets on or off would not be an adequate alternative to our operators, as it might lead to constraint violations:
For example, a feature $f_1$ in the atomic set might require another feature $f_2$ which is not part of the atomic set, as it does not require $f_1$ as well. Lienhardt et al. \cite{lienhardt2020lazy} evaluated the benefits of using \emph{feature model interfaces} \cite{schroter2016fminterfaces} in the performance of valid configurations discovery. Slicing the feature model in feature model fragments, that hide some of the features and dependencies, can be a direction of future work to further optimize our approach.

\mypar{Prioritizing of user needs}
Configuration processes usually involve several stakeholders with their own needs and non-functional expectations.
To support this, staged configurations of feature models~\cite{stagedConfiguration}, multi-views with configuration work flows~\cite{confworkflows}, and one-dimensional approximations \cite{peng2021veer} have been proposed.
In our work, we focus on complete automatic configuration based on non-functional properties, and provide a set of solutions with their trade-offs.
This allows stakeholders to make their own choice without requiring stakeholder preferences to be captured and encoded; a difficult and fragile process.

\mypar{Consistency-preserving model transformation}
In a line of research on combining model transformation with search-based software engineering \cite{hegedus2015model,fleck2016search,Burdusel2018mdeo}, several works deal with validity during generation and analysis of transformation rules.
Kosiol et al.~\cite{Kosiol+22} support a consistency notion and associated analysis to reason about the validity impact of a particular rule; however, they do not consider rule generation.
Kehrer et al.~\cite{kehrer2016automatically} introduced an approach for generating sound and complete set of edit rules for a given metamodel.
The supported notion of validity is focused on a certain type of multiplicity constraints (closed multiplicities on both sides of a reference).
Burdusel et al.~\cite{burdusel2019automatic,burdusel2017towards} generalized this approach to support arbitrary multiplicity constraints, albeit without a completeness guarantee.
They also applied the generated rules to search-based optimization, in the context of their MDEOptimiser tool~\cite{Burdusel2018mdeo}.
For our considered problem, we cannot benefit from these earlier works:
Since they are only tailored towards simple metamodels and not towards feature models with their complex configuration spaces, they do not include any means for addressing the combinatorial effects that we tackle with our technique. 

	


 \section{Conclusions}
\label{section:conclusion}

We have introduced consistency-preserving configuration operators (CPCOs), which capture the changes required for a feature configuration to maintain consistency whenever a feature is activated or deactivated. 
We have shown that CPCOs are useful for improving the search for optimal configurations of product lines whose features are annotated with additional quality information. 
CPCOs enable the search to converge significantly faster than standard mutation operators used in state-of-the-art genetic algorithms.
While generating CPCOs introduces a certain performance overhead, that overhead becomes less important in scenarios where the optimal configuration changes over time (e.g., because of attribute value changes), in which the generated CPCOs can be reused at no extra cost.
Our work also helps address a general lack of cross-tool comparisons in the SPL field.
Our evaluation-experiment infrastructure should be useful for other researchers attempting cross-tool comparison of optimization approaches.
Providing such a common interface and dataset to allow comparison of SPL optimization tools is a key technical challenge in the community~\cite{struber2019facing}.

We envision the following directions of future work:
First, we plan to broaden the scope of our experiments, especially to also take into account feature interactions, different distributions of attribute values, and tool parametrizations.
Second, 
we aim to further improve the efficiency of crossover by avoiding rule sequences in which previous configuration decisions in a sequence of rule applications are reverted.
This requires that the CPCO sequences are non-conflicting, which can be determined using an existing efficient static analysis \cite{lambers2018multi}.
This analysis should also get easier because of the specialised nature of our rules.

Finally, we foresee additional use cases for CPCOs beyond optimisation:
CPCOs might facilitate formal analysis of product lines \cite{thum2014classification} by enabling a more efficient exploration of the variant space that focuses on valid variants.
CPCOs can also be used to easily implement configuration editors that ensure any (partial) configuration selected is valid by construction (e.g., for staged configuration~\cite{stagedConfiguration}). 
Both use-cases would benefit from the soundness guarantee introduced in the present work.
That way, CPCOs pave the way for a variety of new research efforts in product line engineering.



%




  \section*{Acknowledgment}

%
The work of Jose-Miguel Horcas was supported by the Spanish SRUK/CERU International Mobility Programme (On the Move) 2018/2019, and the projects \emph{MEDEA} RTI2018-099213-B-I00 (co-financed by FEDER funds), \emph{Rhea} P18-FR-1081 (MCI/AEI/FEDER, UE), \emph{LEIA} UMA18-FEDERIA-157, \emph{DAEMON} H2020-101017109, and \emph{OPHELIA} RTI2018-101204-B-C22.
The work of Daniel Strüber was partially supported by the Deutsche Forschungsgemeinschaft (DFG), grant 413074939.
The work of Alexandru Burdusel was supported by the EPSRC with award reference 1805606.
We would like to thank Jens Kosiol for his comments on an earlier draft.

\bibliographystyle{IEEEtran}
\bibliography{IEEEabrv,00_references}

%

\newenvironment{biography}[2]
  {\begin{IEEEbiography}[{\includegraphics[width=1in,height=1.25in,clip,keepaspectratio]{#2}}]{#1}}
  {\end{IEEEbiography}}

\newpage

\begin{biography}{José Miguel Horcas}{horcas}
is a postdoc researcher at the University of Málaga, Spain, where he received his PhD in Computer Sciences in 2018. His main research areas are related to software product lines, including variability and configurability, and quality attributes. He carried out a postdoc stay at King’s College London, UK, in 2019 within the SRUK/CERU International Mobility Programme (On the Move). More information available at \url{https://sites.google.com/view/josemiguelhorcas}.
\end{biography}

\begin{biography}{Daniel Str\"uber}{strueber}
	is a senior lecturer at Chalmers | University Gothenburg, Sweden, and an assistant professor at 
Radboud University Nijmegen, The Netherlands. His research is in model-driven engineering, software product lines, and AI engineering. He has co-authored over 75 papers and is the project lead of Henshin, a model transformation language used in academia and industry in more than 15 countries.
Papers and more information are available at \url{www.danielstrueber.de}.
\end{biography}

\begin{biography}{Alexandru Burdusel}{burdusel}
  received his PhD degree in computer science from King’s College London in 2021. Previously he worked as a software engineer in industry. His current research interests are in optimisation methods, model-driven engineering and search-based software engineering.
\end{biography}

\begin{biography}{Jabier Martinez}{martinez} is a research engineer in the Digital Trust Technologies (TRUSTECH) area of Tecnalia since 2018. His background is on providing methods and tools for systems modelling and variability management. After several years of industrial experience, he received his PhD from the Luxembourg and Sorbonne Universities with an awarded thesis about product line adoption and analysis. He co-organizes the Reverse Variability Engineering workshops. His interests also include non-functional properties.
\end{biography}

\begin{biography}{Steffen Zschaler}{zschaler}
  is Reader in Software Engineering at King's College London, UK. His research focuses on the foundations, tools, and applications of model-driven engineering, including search-based approaches to finding optimal models (through the MDEOptimiser tool) and modelling languages for variability management (e.g., VML*). He obtained his doctoral degree from Technische Universit\"at Dresden, Germany. More information is available at \url{www.steffen-zschaler.de}.
\end{biography}

\vspace{40pt}

\clearpage{}
\appendices

\section{CPCO Generation}
\label{sec:appendix:cpco_gen}

  In this appendix, we provide  detailed information about our activities for \textit{discarding unnecessary VB-rule instances} (gray parts of Fig.~\figref{activity_diagram}; described on a high level as part of Sect.~\ref{section:generating_cpcos:analysing_fasds}).
	This complements our presentation of the other steps (white parts of Fig.~\figref{activity_diagram}), which are presented in detail in Sect.~\ref{section:generating_cpcos:analysing_fasds}.
  
  Specifically, we apply the following activities:

  \begin{enumerate}
    \item \emph{Constraints for or-overlaps.} 
          Consider the excerpt of the feature-activation diagram from Fig.~\figref{fad-example} that's shown in Fig.~\figref{or_overlap_example}.
          Note how many or-nodes have paths that lead to the same feature decisions.
          For example, O$_{61}$, O$_{71}$, O$_{01}$ all lead to (F5$^+$). 
          This information isn't captured in the VB-rule we are currently generating and, as a result, the current VB-rule feature model allows multiple configurations that lead to the same generated rule.
          For example, selecting O$_{61}$ and O$_{02}$ selects the same set of feature decisions as selecting O$_{62}$ and O$_{01}$.
          In addition to producing such redundant rule instances, we are also generating unnecessarily large repairs for a feature decision. 
          For example, if OR$_0$ is reached via O$_{62}$, then we have already made the decision to activate F6. 
          Also activating F5 does not improve the repair for the deactivation of F7 (which ultimately triggered OR$_0$), it just makes our operator larger---potentially a lot larger depending on the consequences of activating F5.
          We avoid such situations by adding explicit constraints that correlate decisions by different or-nodes.
          Specifically, for each direct follow-node of any or alternative that can also be reached on another path, we generate a condition that ties the decisions on both paths together.
          In our example, we would, for example generate a condition $O_{61} \wedge O_0 \Rightarrow O_{01}$ stating that if we have selected O$_{61}$ and we have to make a decision about O$_0$, we will always choose O$_{01}$.
					
    %
    \item \emph{Blocking of self-activating cycles.} 
          Figure~\figref{or_overlap_example} also demonstrates another problem: self-activating cycles.
          In generating the VB-rule feature model, we have been able to avoid having to enumerate all repair paths by only encoding direct implications between or-alternatives and the directly following or-nodes and relying on the transitive nature of the logical implication to correctly reconstruct the paths on rule instantiation.
          However, this construction allows some superfluous rule instances to be constructed, too, namely, where there are cycles in the or-implications.
          For example, O$_{22}$ requires a decision to be made about OR$_6$ and, in turn, O$_{62}$ requires a decision about OR$_2$. 
          With the VB-rule feature model so far, activating or-alternatives in a cycle is always possible, even without a path from the root decision.
          This produces unnecessary repairs, making the operator unnecessarily complex.
          Because there are many cycles in a feature-activation diagram, and every cycle can be activated freely, we end up producing an unnecessarily large number of rule instances.
          This can be fixed if we can add constraints to ensure cycles can only be activated if there is a path from the root decision into the cycle.
          
          Collecting all cycles in a directed graph is computationally complex~\cite{Johnson1975}.
          However, we do not actually need to collect all nodes of all cycles. 
          It is sufficient for us to identify one link in each cycle to be broken unless a path into the cycle is also active.
          It is enough for the entering path to lead into the cycle from outside, we do not have to check whether it starts at the root decision; this will be taken care of by the transitive nature of the constraint we are adding.
          
          Algorithm~\ref{alg:or-implication-cycles} (and Algorithm~\ref{alg:effective-set}) shows how we can use a variation of the standard depth-first approach to cycle breaking (searching for back-edges) in directed graphs to collect the information we require.
          After running the algorithm over the graph of or-implications (we remove the specific feature-decisions for this analysis to improve the efficiency of the depth-first search), we generate a constraint for each breaking or-alternative requiring that a cycle-entering path must also be active.
          
          \begin{algorithm}[tbp]
            \caption{Breaking cycles in the or-implication graph (a.k.a.  \textit{Or-Cycles} in Figure~\figref{activity_diagram}).}\label{alg:or-implication-cycles}
            \begin{algorithmic}[1]
              \scriptsize
              
              \Require{\textit{rootFeature}: the root feature of the VB-rule.}
              \Ensure{Returns a map from VB-rule features that are on a cycle to sets of features that are entry points to the cycle.
                      For each mapping $f \mapsto s$ in this map, we will generate constraints that require for $f$ to be activated at least one feature in $s$ to be activated.
                      As a result, cycles can only be activated if a path from outside the cycle (and, thus, from the root feature) has been activated.}
              \Function{computeCycles()}{}
                \State $\textit{visited} \gets \emptyset$ \Comment Set of nodes visited.
                \State $\textit{stack} \gets \emptyset$ \Comment Stack of nodes visited in current traversal branch.
                \State $\textit{cycleEntries} \gets \emptyset$ \Comment Map from features to sets of (add,delete) tuples. 
                \Statex                
                \State \Call{recursivelyComputeCycle}{$\textit{rootFeature}, \textit{stack}, \textit{visited}, \textit{null}$}
                \Statex
                \LineComment Resolve cycle entry data. \textsc{mapValues} takes a map and produces a new map with the given function applied to each value.
                \State \Return $\textit{cycleEntries}.\textsc{mapValues}[\textsc{effectiveSet}]$\label{alg:computeCycles:effectiveSet}
              \EndFunction
              \Statex
              \Require{\textit{addDeleteTupleSet}: a set of tuples. 
                       The two elements of each tuple represent or-nodes or or-alternatives to be added to, and deleted from, the overall set, respectively.
                       These two elements of each tuple can be accessed through projection functions written below using standard OO dot notation as $x.add$ and $x.delete$, respectively.}
              \Ensure{Returns a set of or-nodes and or-alternatives. 
                      This is the union of all added elements minus the union of all deleted elements.}
              \Function{effectiveSet}{addDeleteTupleSet}
                \State \Return $\left\{a | x \in addDeleteTupleSet, a = x.add\right\} \setminus$ \Statex $\qquad\qquad\left\{d | x \in addDeleteTupleSet, d = x.delete\right\}$
              \EndFunction
            \end{algorithmic}
          \end{algorithm}
          \begin{algorithm}[tbp]
            \caption{Computing the effective set of cycle entries.
                     For compactness, we use += and -=, respectively, to add and remove elements to/from sets and maps. 
                    Individual mappings in a map are represented using the standard $\mapsto$ operator.}
            \label{alg:effective-set}
            \begin{algorithmic}[1]
              \scriptsize
              
              \Require{\textit{feature}: the VB-rule feature being visited.}
              \Ensure{\textit{stack} tracks or-features on the path from root that we are currently on.}
              \Ensure{\textit{visited} globally tracks visited VB-rule features.}
              \Ensure{\textit{comingFrom} tracks the or-alternative visited directly before, if any.}
              \Function{recursivelyComputeCycle}{\textit{feature}, \textit{stack}, \textit{visited}, \textit{comingFrom}}
                \If{$\textit{feature} \in \textit{visited}$} 
                  \If{\textit{feature} is `root' or \textit{feature} is or-alternative}
                    \State \Return $\emptyset$ 
                  \Else
                    \Comment \textit{feature} is an or-feature
                    \If{$\textit{feature} \in \textit{stack}$}\label{alg:computeCycles:feature-in-stack} \Comment We've found a cycle
                      \LineComment \textit{preAlternatives} is a map from or-features to all the or-alternatives that imply them in the VB-rule (the inversion of the edges in the or-implication graph).
                      \State $\textit{cycleEntries}$ += $\textit{feature} \mapsto$ \Statex $\qquad \qquad \qquad \qquad \left(\textit{add}: \textit{preAlternatives}[\textit{feature}], \textit{del}: \textit{comingFrom}\right)$
                      \State \Return $\left\{\textit{feature}\right\}$
                    \Else
                      \State \Return $\emptyset$
                    \EndIf
                  \EndIf
                \EndIf
                \Statex
                \State \textit{visited} +=\textit{feature}
                \If{\textit{feature} is or-alternative}
                    \LineComment Remove \textit{feature} from cycle entries for any or-feature on \textit{stack}
                    \LineComment for which we have already recorded a cycle.
                    \ForEach{$\textit{of} \in \left\{\textit{orF} \in \textit{stack}~|~\textit{orF}\mapsto X \in \textit{cycleEntries}\right\}$}\label{alg:computeCycles:pre-stepdown-start}
                      \State \textit{cycleEntries} += $\textit{orF} \mapsto \left(\textit{add}: \emptyset, \textit{del}: \textit{feature}\right)$
                    \EndFor\label{alg:computeCycles:pre-stepdown-end}
                \ElsIf{\textit{feature} is or-feature}
                  \State \textit{stack} += \textit{feature}
                \EndIf
                \Statex
                \LineComment Step down
                \State $\textit{result} \gets \textit{feature}.\textit{edges}.\textsc{flatMap}[$
                \State $\qquad \Call{recursivelyComputeCycles}{\textit{stack}, \textit{visited}, \textit{feature}}]$
                \Statex
                \If{\textit{feature} is or-feature}\label{alg:computeCycles:post-stepdown-start}
                  \State \textit{stack} -= \textit{feature}
                  \LineComment Update incoming edges
                  \If{$\textit{feature} \mapsto X \in \textit{cycleEntries}$}\label{alg:computeCycles:start-of-cycle}
                    \LineComment \textit{cycleEntries} contains at least one mapping for \textit{feature}, which means we have encountered a cycle containing and to be broken by \textit{feature}.
                    \LineComment Ensure we add incoming edges only for features we have visited except for \textit{feature}.
                    \State $\textit{result} \gets \textit{result} \setminus \left\{\textit{feature}\right\}$ \Comment{Also mark cycle completed}
                    \ForEach{$f \in \textit{result}$}
                      \State \textit{cycleEntries} += $f \mapsto X$
                      \State \textit{cycleEntries} += $f \mapsto \left\{\textit{add}: \emptyset, \textit{del}: \textit{comingFrom}\right\}$
                    \EndFor
                  \Else
                    \LineComment \textit{feature} isn't involved in any cycles as an "endpoint", so all entries, except the one we came in on, need to be added as potential entry points for any cycle we have found during descent.
                    \LineComment At this point result cannot contain \textit{feature}, so there is no need to remove it.
                    \ForEach{$f \in \textit{result}$}
                      \State \textit{cycleEntries} += $f \mapsto$ \State $\qquad \left(\textit{add}: \textit{preAlternatives}[\textit{feature}], \textit{del}: \textit{comingFrom}\right)$
                    \EndFor
                  \EndIf
                \EndIf\label{alg:computeCycles:post-stepdown-end}
                \Statex
                \State \Return \textit{result}
              \EndFunction
            \end{algorithmic}
          \end{algorithm}

          To understand how the algorithm works, let us consider the example or-implication graph in Fig.~\figref{or-cycle-example}.
          It contains 4 cycles (labelled I to IV), none of which contains the `root' feature of the VB rule.
          VB-rule features in each cycle should only be allowed to be activated if one of the features connecting the cycle to the root feature has been activated.
          For example, the two features in Cycle~I should only be activated if $OA_3$ has also been activated.
          Similarly, Cycle~II should only be activated if $OA_3$, $OA_8$, or $OA_9$ have been activated.
          
          \insertFigure[onecolumn=T,caption={Example or-implication graph with 4 cycles}]{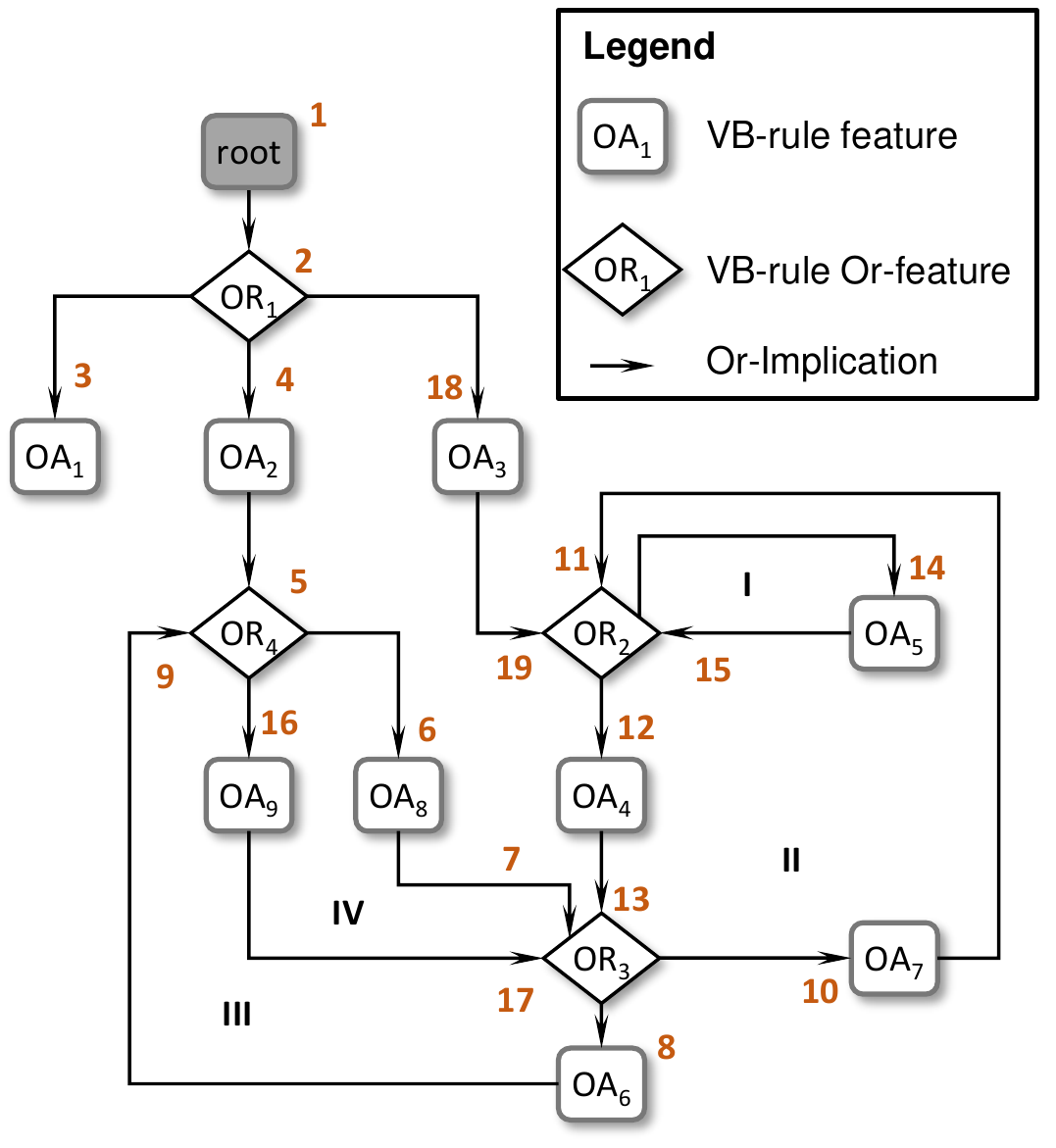}

          The Arabic numerals in Fig.~\figref{or-cycle-example} show one possible sequence in which the nodes in the or-implication graph might be visited by a depth-first search.
          Visits~1 to 8 are fairly straightforward and do not trigger any special conditions.
          As $OR_4$ is visited in Step~9, $\textit{stack} = \left(OR_1, OR_4, OR_3\right)$ and $\textit{comingFrom} = OA_6$.
          This triggers the condition in Line~\ref{alg:computeCycles:feature-in-stack} of Algorithm~\ref{alg:or-implication-cycles}.
          We note the fact that we have found a cycle and will break it at feature $OR_4$ by adding an entry to the global \textit{cycleEntries} map.
          This entry states that $OR_4$ should be activated when any of the or-alternatives pointing at it (to this end, \textit{preAlternatives} is the inversion of the edges in the or-implication graph) \emph{except} for $OA_6$, which we know to be part of the cycle because that was the way we reached the current node.
          In terms of the classic depth-first search algorithm for breaking cycles, we will eventually break the edge between $OA_6$ and $OR_4$ by making the activation of the or-node dependent on a path from `root' being active, too.
          Note that we have to keep activating and deactivating entry points separate to handle over-lapping cycles. 
          When we calculate the final set of cycle entry points using \textsc{effectiveSet} (Line~\ref{alg:computeCycles:effectiveSet}), we will only consider nodes that are in the \textit{add} set but not in the accumulated \textit{delete} set.
          
          An example of the need for this can be seen as the algorithm progresses.
          First, we backtrack to $OR_3$, where we find another alternative to explore. 
          This eventually leads us to a similar place when we visit $OR_3$ again in Step~13.
          Here, $\textit{stack} = \left(OR_1, OR_4, OR_3, OR_2\right)$ and $\textit{comingFrom} = OA_4$.
          We record the new cycle by adding an entry to \textit{cycleEntries} for $OR_3$ allowing activation when $OA_9$ or $OA_8$, but not $OA_4$ have been activated.
          Next we back track to $OR_2$, where we explore the second alternative (Steps~14 and 15) and find a nested cycle.
          When we visit $OR_2$ again in Step~15, $\textit{stack} = \left(OR_1, OR_4, OR_3, OR_2\right)$ and $\textit{comingFrom} = OA_5$.
          We record the new cycle by adding an entry to \textit{cycleEntries} for $OR_2$ allowing activation when $OA_7$ or $OA_3$, but not $OA_5$ have been activated.
          We backtrack to the beginning of Cycle~I and enter the code on Lines~\ref{alg:computeCycles:post-stepdown-start}--\ref{alg:computeCycles:post-stepdown-end}.
          This code is needed because cycles may have entry points at any node along the way (it is enough to track entries into or-nodes as every or-alternative will have to have a preceding or-node).
          For example, we need to record the fact that Cycle~II can be reached via $OA_3$.
          We do this by adding all entry paths for every or-node to all cycles that we are currently working back from (because these are the end points of cycles for which we have yet to back track to the beginning of the cycle again).
          We distinguish two situations (see the condition on Line~\ref{alg:computeCycles:start-of-cycle}): 
          Where the current or-node is also the ``end point'' of a cycle we make sure to only add those paths that aren't part of any cycle found so far.
          Otherwise, we add all incoming paths.
          In either case, we explicitly disallow activation via the or-alternative through which we have reached the current or-node (because we know we are in a larger cycle).
          In our case, this asserts that Cycle~II may be activated by $OA_3$ in addition to what we have recorded previously in Step 13.
          This correctly collects all entry points to Cycle~II.
          We are not adding any additional entry points to Cycle~I because we recognised that we have found the starting point for this cycle and, as a result, have removed $OR_2$ from \textit{result}.
          
          We need to consider one final scenario:
          When visiting $OR_3$ in Step~17, we immediately return because we have already visited this node before.
          As a result, we never fully explore Cycle~III and would not record the correct way of breaking this cycle.
          This is good because it keeps the computational complexity of the algorithm to the standard complexity of depth-first search, but we need a way of recording the fact that $OA_9$ should not be allowed to activate $OR_4$.
          The code on Lines~\ref{alg:computeCycles:pre-stepdown-start}--\ref{alg:computeCycles:pre-stepdown-end} does exactly that.
          As we go forward through the graph, at each or-alternative we visit, we check the or-nodes that we have visited so far (\emph{i.e.,} they are in \textit{stack}) and for which we have identified a cycle at some previous point (\emph{i.e.,} they already have an entry in \textit{cycleEntries}).
          At Step 16, this is precisely $\left\{OR_4\right\}$.
          For each such or-node, we add an entry to \textit{cycleEntries} to exclude the current or-alternative.
          As a result, in Step 16 we add an entry forbidding $OR_4$ to be activated by an activation of $OA_9$ alone, which effectively breaks Cycle~III.
                    
    \item \emph{Removal of dead VB-rule features.} 
          We identify dead VB-rule features and remove them from the VB-rule.
          As a result, some feature decisions will have an empty presence condition, indicating they can never be part of any instance of the VB-rule. 
          We remove these feature decisions from the VB-rule.
  \end{enumerate}  
	
	\section{Soundness argumentation}
	\label{sec:appendix:soundness}

  In this appendix, we provide a detailed and precise argumentation for the soundness of our CPCO generation algorithm.
	We provide three theorems corresponding to activities shown in the overview in Fig.~\figref{activity_diagram}, and explained in Sect.~\ref{section:generating_cpcos} and the previous appendix:
	Theorem 1 focuses on Algorithm~1, Theorem 2 addresses Algorithm~2, and Theorem~3 deals with our measures for removing unnecessary CPCO instances (gray parts of the figure, including Algorithms 3 and 4 from the previous appendix).
  
We first provide a definition of variability-based rules, our chosen representation of CPCOs.
Compared to previous work \cite{struber2018variability}, our definition is simplified, as it precisely matches the structure of rules that we generate as part of our approach.
Where necessary, we refer back to previous definitions from the main paper body.

\begin{definition}[Rules and VB rules]\label{def:vbrule}
A rule $r = (N_r, a_r)$ consists of a set of nodes $N_r$ and a function $a_r : N_r \rightarrow (String, Option(\mathbb{B}), \mathbb{B})$ that maps each node from $N_r$ to an attribute value change $a = (\textit{name}_a, \textit{old}_a, \textit{new}_a)$.
The attribute name $\textit{name}_a$ and the new value $\textit{new}_a$ must be non-null, whereas the old value $\textit{old}_a$ may be null.
\\
A variability-based rule (VB rule) $\hat{r} = (r_{\hat{r}}, \mathcal{F}_{\hat{r}}, pc_{\hat{r}})$ consists of a rule $r_{\hat{r}}$, a feature model $\mathcal{F}_{\hat{r}}$ (Def.~\ref{def:fm}), and a function $\textit{pc}_{\hat{r}} : N_r \rightarrow \textit{Bool}(F_{\hat{r}})$ that maps each node in $N_r$ to a propositional formula over features from $\mathcal{F}_{\hat{r}}$.\\
A configuration $c$ of $\mathcal{F}_{\hat{r}}$ induces a rule $r$, called \textit{flattened rule}, that is obtained by replacing in all presence conditions the feature names with the values from $c$, and removing nodes whose presence condition evaluates to \textit{false}.\\
The set $\textit{Flat}(\hat{r})$ is the set of all flattened rules that can be obtained by possible configurations of a VB rule $\hat{r}$.
\end{definition}

For example, Fig.~\ref{fig:VBruleInstance} shows two rules, both with three nodes, each of which contain one attribute value change.
In both rules, a node labeled \textit{Screen3} specifies the value of the  attribute \textit{active} to be changed from \textit{true} to \textit{false}, whereas the other nodes just specify a new value.
Together, both rules form the set $\textit{Flat}(\hat{r})$ for the rule shown in Fig.~\ref{fig:ActSMSTransfer}, which shows the feature model and the PCs (in gray, dashed boxes) as part of the figure. 

To reason about the semantics of changes expressed as VB rules, we now define the notion of toggle graph.
Recall the term \textit{feature decision}, which refers to one individual decision to either activate or deactivate a specific feature.
Toggle graphs capture our intuition of \textit{paths}---a set of feature decisions that, if executed in concert, preserve validity.

\begin{definition}[Toggle graph]\label{def:togglegraph}
Let a feature activation diagram $G_d = (V_d,E_d)$ (Def.~\ref{def:fad}) and a feature decision $f_{!}$ for a feature $f \in F$ be given.
A \emph{toggle graph} for $f_{!}$,  written $G_{f_{!}} = (V_{f_{!}},E_{f_{!}})$, is a subgraph of $G_d$ with the following properties:\\
The edge set $E_{f_{!}}$ is a subset of $E_d$ where
\begin{enumerate}
	\item $E_{f_{!}}$ contains all outgoing edges of  $f_{!}$, and 
	\item for each edge $e \in E_{f_{!}}$, the following applies:
if the target node of e, written $\textit{trg}(e)$, is a feature decision, $E_{f_{!}}$ contains all outgoing edges of  $\textit{trg}(e)$; else, $\textit{trg}(e)$ is an or-node and $E_{f_{!}}$ contains exactly one outgoing edge of  $\textit{trg}(e)$.
\end{enumerate}
The vertex set $V_{f_{!}}$  is the union of the source and target nodes of the edges in $E_{f_{!}}$.\\
A toggle graph is \emph{valid} if it contains at most one feature decision for every feature $f \in F$.
A toggle graph is \emph{applied} to a configuration $c$ by applying all of its feature decisions to $c$.
\end{definition}

In general, there are multiple toggle graphs for a particular feature decision, since there are multiple ways to choose the outgoing edge for or-nodes.
For an example, consider the feature activation diagram in Fig.~\figref{fad-example}, and the explanations of the green and orange paths (a.k.a. toggle graphs) in the text.


The following theorem shows the soundness of Algorithm~1, by ensuring that the toggle graphs that can be derived from the generated feature activation diagrams indeed represent sets of changes that together preserve validity.

\begin{theorem}[Soundness of Algorithm 1]
Let the following be given:
a feature model $\mathcal{F}$, a configuration $c \in \textit{cfg}(\mathcal{F})$, and the feature activation diagram $G_d$ generated by Algorithm 1. 
Applying a valid  toggle graph $G_{f_{!}}$ of $G_d$ to $c$ leads to a valid configuration $c' \in \textit{cfg}(\mathcal{F})$.
\end{theorem}

\textbf{Proof sketch}: Applying a toggle graph $G_{f_{!}}$ to c as per Def.~\ref{def:togglegraph} yields another configuration. Let us call this configuration~$c'$.
The validity of $c'$ follows from the application of principles during the generation of the feature activation diagram (line 7 in Algorithm 1) and from the definition of toggle graph (Def.~\ref{def:togglegraph}), which encodes the idea of recursively including consequences of feature decisions. In more detail:

For all feature model constraints, we show that they are still fulfilled after the activation and deactivation of particular features.
Let $f$ be a feature that is activated by $G_{f_{!}}$.

\mysubparproof{\cmand, \cpar, \creq} These constraints require a particular feature to be activated if $f$ is activated, i.e., all mandatory children, parent features, and required features of $f$. They are addressed by the principles \actmand, \actpar\ and \actreq: the feature activation diagram contains an activation vertex for all such features.
From Def.~\ref{def:togglegraph}, we know that if the activation of $f$ is included, these activation vertices are included as well, ensuring that the constraints are satisfied.

\mysubparproof{\cor, \cxor} These constraints require one sub-feature to be activated if a group feature $f$ is activated. 
 For each activation of a ``or'' and ``xor'' group, the feature activation diagram contains an ``or'' node, via application of \actgroup. 
Per definition, the toggle graph contains an outgoing edge for the ``or'' node, and contains the target feature decision of that edge.
That leads to the activation of a feature $g$ in the graph, and hence, to the satisfaction of \cor\ in $c'$.
For \cxor, we need to ensure that $g$ is the only feature activated in this group.
This is the case because applying \actxor\ ensures that all sibling features of $g$ will be deactivated.
By definition, the relevant decision is also contained in the toggle graph.

\mysubparproof{\cexcl}  This constraint requires certain features to be deactivated if $f$ is activated, which is ensured via \actexcl, also included into the toggle graph.

\mysubparproof{\croot} This constraint requires that the root is always activated. In $c'$, this is the case because only real-optional features are deactivated or activated via principle applications.

\newcommand*{\QEDA}{\hfill\ensuremath{\blacksquare}}
For the deactivation of a feature f, the argumentation for the constraints is completely dual. $\QEDA$
\medskip

The following theorem and corollary show the soundness of Algorithm 2, by ensuring that the information represented in a generated CPCO is a set of changes that together preserve validity (a.k.a. a toggle graph).

\begin{theorem}[CPCO variants represent valid toggle graphs]
Let the following be given:
a feature model $\mathcal{F}$,  the feature activation diagram $G_d = (V_d,E_d)$ generated by Algorithm~1, a feature decision~$f_!$, 
and the CPCO $c_{f_{!}}$ generated for $f_!$ using the basic CPCO generation algorithm (white parts of Fig.~\figref{activity_diagram}, including Algorithm~2).
For every CPCO variant $r \in Flat(c_{f_{!}})$, there exists a valid toggle graph $G_{f_{!}} \subseteq G_d$  with the same feature decisions as $r$.
\end{theorem}

\textbf{Proof sketch}:  We argue over the structure of the generated CPCOs, which are generated in the form of VB rules (Def.~\ref{def:vbrule}, based on the information collected in Algorithm 2 (see the one shown in Fig.~\figref{detailed_vb_rule_example}).
Recall the following components:
(i) Each rule node generated corresponds to a feature decision $f_! \in V_d$.
For simplicity, we refer to a rule node and the corresponding feature decision as the same entity (for full accuracy, one could define a mapping function).
(ii) Each rule node $f_!$ is annotated with presence conditions, written  $\textit{pc}(f_!)$, that arises from a disjunction over rule feature names---specifically, the \textit{root} feature and or-group alternatives--which represent different other nodes from which this node can be reached.
The rule feature model consists of:
(iii) a root feature with several or-group children, each of which with multiple alternatives representing choices of switching one out of several features on; 
(iv) implications that represent the transition from one node to another, and
(v) additional constraints enforcing that there cannot be an activation and deactivation node for the same feature.

A CPCO rule variant $r$ is one rule obtained by configuring a VB rule $\hat{r}$.
Given a particular CPCO rule variant $r$, we have to show the existence of a toggle graph with the same feature decision nodes as  $r$. As a candidate for this toggle graph, we define a graph $G_{f_{!}} = (V_{f_!}, E_{f_!})$ as follows:
\[ V_{f_!} = N_r~\cup~\textit{orSucc} \]
\[ E_{f_!} = \{e \in E_d | src(e), trg(e) \in V_{f_!}\} \]
\[ \textit{orSucc} = \{o \in V_d~|~o~\textit{is}~\textit{an}~\textit{or-node},\] \vspace{-15pt}
\[	\exists (e,n) \in (E_d,N_r)~s.t.~src(e) = n, trg(n) = o\}   \]

That is, the vertex set $V_{f_!}$ is the union of the set $N_r$ of feature decisions~from~$r$ and the set \textit{orSucc} of or-nodes from $G_d$ succeeding any of these feature decisions.
The edge set $E_{f_!}$ consists of all edges from $G_d$ that connect two of $V_{f_!}$'s elements.
 
We need to show that $G_{f_{!}}$ is a valid toggle graph, as per Def.~\ref{def:togglegraph}.
To this end, we argue over the structure of the generated rules.

\smallskip
\textit{Def. ~\ref{def:togglegraph}, condition 1:} 
To show that all outgoing edges of the feature decision $f_!$ are in $G_{f_{!}}$, we first show that $f_!$ itself is in $G_{f_{!}}$.
That is the case because, due to component (ii), the presence condition $\textit{pc}(f_!)$  is a disjunction that, as one clause, includes the feature \textit{root}.
Per component (iii), the feature \textit{root} is mandatory, rendering $\textit{pc}(f_!)$   \textit{true} in all configurations.
Consequently, $f_!$ is contained in any of the rules in $Flat(c_{f_{!}})$, including $r$.

We need to show that all direct successors of $f_!$---i.e., all nodes reachable via an outgoing edge from $f_!$---are included in $V_{f_!}$.
Consider one such node $g \in \{n \in N_d| \exists e \in E_d~s.t.~\textit{src}(e) = f_!, \textit{trg}(e) = n\}$.
If $g$ is a feature decision, consider that the VB rule $\hat{r}$ includes all nodes reached in the traversal of Algorithm 2 (recursive call in line 12), including $g$.
 Moreover, $g$ has a presence condition, which arises from a disjunction that includes $f_!$'s  presence condition (line 5), including the $root$ feature.
Recall that $root$ is always active, and hence, $g$ is always included in $V_{f_!}$. 
If $g$ is an or-node, $g \in G_{f_!}$ holds via the definition of $V_{f_!}$.
Finally, via the definition of  $E_{f_!}$, all edges connecting $f_!$ to one of its successors are included in $E_{f_!}$, leading to  condition 1 being fulfilled.



\smallskip
\textit{Def. ~\ref{def:togglegraph}, condition 2:} 
For a given edge $e \in E_{f_!}$, we consider the target node $g := \textit{trg}(e)$.
Via $E_{f_!}$'s definition, $g$ is in $V_{f_!}$.

If $g$ is a decision node, we need to show that all outgoing edges of $g$ are contained in $E_{f_!}$.
Let us consider any such edge $e' \in \{e \in E_d | \textit{src}(e) = g\}$.
The target of $e'$ can be an or-node or a feature decision.
If it is an or-node, the definition of $G_{f_!}$ ensures that $e'$ is contained in $E_{f_!}$.
Else, we argue in the same way as for condition (i), except for the details of the presence-condition handling. 
Specifically, Alg.~2 propagates the presence conditions of feature decisions to direct successor feature decisions via a disjunction (line 5).
Therefore,  the presence condition of $g$ is stricter than the presence condition of all of its direct successor feature decisions, and all CPCO variants that contain $g$ also contain all of its successor feature decisions.
Hence, direct successors of $g$ (and, via the definition of, $E_{f_!}$, the edge that connects them) are also contained in  $N_{f_!}$ (and $E_{f_!}$).

Otherwise, if $g$ is an or-node, we need to show that exactly one outgoing edge of $g$ is contained in $E_{f_!}$.
For or-nodes, Alg.~2 creates an xor-group feature  $\textit{OR}_{i}$ (line 20), and, for each edge leaving the or-node, a feature $O_{ij}$ (lines 21-22) which becomes a child feature of $\textit{OR}_{i}$ in component (iii).
In addition, Alg.~2 collects a map of feature decisions to succeeding or-nodes (line 23), which is used in component (iv) to generate constraints of the form ``$\textit{pc}(e_{g_{in}}) \implies \textit{OR}_{i}$'', meaning that 
the presence condition of the edge from which we arrive at the or-node implies that the xor-group generated for the or-node has to be activated.
Therefore, if $g$ is included in $V_{f_!}$, $O_i$ is activated, and exactly one of the edges leaving $g$ is included in $E_{f_!}$, leading to condition 2 being fulfilled.



\smallskip
\textit{Def. ~\ref{def:togglegraph}, condition for validity:} 
Since conditions 1  and 2 are fulfilled, $G_{f_!}$ is a toggle graph.
It remains to be shown that it is a valid one.
This is the case because, due to the constraints generated in component (v), there is one constraint for each pair of activation and deactivation nodes of the form ``$\textit{pc}_{f_-}\implies~!\textit{pc}_{f_+}$'', ensuring that both nodes can never be included in the same CPCO variant and hence, also not in the toggle graph $G_{f_!}$. As a result, $G_{f_!}$ is valid.$\QEDA$

\begin{corollary}[Soundness of Algorithm 2]
Applying a CPCO variant from a CPCO generated by
the basic CPCO generation algorithm (white parts of Fig.~\figref{activity_diagram}, including Algorithm~2)
 to a valid configuration yields a valid configuration again.
\end{corollary}

\textbf{Proof sketch}: This corollary follows directly from Theorems 1 and 2.

Please note that we avoid a particular soundness-related complication in Algorithm 2 by representing presence conditions of nodes temporarily as \textit{proxies} that are resolved in a later post-processing step (described in Sect.~\ref{section:generating_cpcos:analysing_fasds}).
Without this post-processing step, we might propagate incomplete presence conditions (e.g., in line 5), since the full presence condition of a node (arising from the different ways in which we can reach it) is generally not known the first time we touch the node.
By instead propagating ``proxy'' presence conditions and resolving them later, we ensure that the definite presence conditions are only set when all information about reachability of nodes is available, leading to consistent information.$\QEDA$

\medskip

The follow theorem shows that our measures to discard unnecessary VB rule instances (simplifications, see gray parts of Fig.~\figref{activity_diagram}) do not threaten the soundness of the overall algorithm.

\begin{theorem}[Valid simplification]
Let the following be given: a feature decision $f_{!}$, 
the CPCO $c_{f_{!}}$ generated by the basic CPCO generation algorithm (white parts of Fig.~\figref{activity_diagram}), and 
the CPCO $c'_{f_{!}}$ generated by the full  algorithm (full Fig.~\figref{activity_diagram}).
The set of CPCO variants
 $\textit{Flat}(c'_{f_{!}})$ is a subset of $\textit{Flat}(c_{f_{!}})$.
\end{theorem}

\textbf{Proof sketch}: 
We consider the three activities that extend the basic generation algorithm:

\smallskip
\textit{(1.) Or-overlaps \& (2.) Or-cycles.}
These two activities alter the generated CPCO exclusively by extending the generated feature model constraints, i.e., conjoining them with additional terms.
In consequence, the altered constraints are stricter than the original ones.
Both activities do not alter the presence conditions in the generated CPCOs.

Making the constraints in a feature model stricter can lead to some configurations being discarded, but it cannot lead to new configurations.
Hence, the following holds for the  sets of configurations:  $\textit{configs}(c'_{f_{!}}) \subseteq \textit{configs}(c_{f_{!}})$, where $c'_{f_{!}}$ indicates the altered CPCO and $c_{f_{!}}$ the original one.
  
CPCO variants are obtained by configuring a  given CPCO.
Since each configuration of $c_{f_{!}}$ is a configuration of $c_{f_{!}}$, and $c'_{f_{!}}$ and $c_{f_{!}}$ are identical except for their feature model constraints, each CPCO variant of $c'_{f_{!}}$ is one of $c_{f_{!}}$ as well.

\smallskip
\textit{(3.) Dead feature removal.}
This activity alters the generated feature sets, constraints and presence conditions in such way that dead features are not included (effectively, replacing them with the value \textit{false}).
Dead features necessarily take on the value \textit{false}, rendering the resulting constraints and presence conditions equivalent to the original ones.
Therefore, the set of CPCO variants that can be generated remains identical.
For determining the set of dead features, we rely on standard dead feature analysis \cite{benavides} via SAT solving, which is known to be sound.$\QEDA$






\begin{corollary}[Soundness of CPCO generation algorithm]
Let a CPCO variant $r$ from a CPCO generated by the full CPCO generation algorithm be given.
Applying $r$ to a valid configuration yields a valid configuration again.
\end{corollary}

\textbf{Proof sketch}: Follows directly from Theorems 3 and Corollary 1.$\QEDA$

\section{Algorithmic Complexity}
\label{section:appendix_complexity}

  A high-level summary of the computational complexity has already been given in the main body of the paper (Sect.~\ref{sec:algo_properties}).
  This appendix provides more detail on the computational complexity of our algorithm.
  We start with an analysis of the computational complexity of the construction of a feature-activation diagram (FAD) in Sect.~\ref{section:appendix_complexity:fad}.
  Constructing CPCOs is based on a traversal of a sub-diagram of the FAD for each real-optional feature, and we analyse the implications of this in Sect.~\ref{section:appendix_complexity:cpco}.

  
  \subsection{FAD construction}
  \label{section:appendix_complexity:fad}
  
    The feature-activation diagram is constructed incrementally, by adding feature decisions and their consequences into an existing feature-activation diagram.
    Where a feature decision already exists in the diagram, no new node is added. Instead, the existing node is reused and its consequences are not explored again.
    As a result, constructing a complete feature-activation diagram is equivalent in complexity to performing a full depth-first search (DFS) over the final feature-activation diagram.
    
    The computational complexity of DFS is $O(N+E)$ where $N$ is the number of nodes in the graph and $E$ is the number of edges in the graph.
    Therefore, we can determine the computational complexity of constructing a feature-activation diagram by asking, for a given feature model, how many nodes and edges does the full feature-activation diagram have?
    We can determine this by analysing the (de)activation principles from Sect.~\ref{section:cpcos} and determining how many nodes and edges each adds to the feature-activation diagram.
    
    Feature-activation diagrams contain two types of nodes: feature-decision nodes and or-nodes.
    For a feature model with $F$ real-optional features, the full feature-activation diagram will contain $2F$ feature-decision nodes (for each feature an activation and a deactivation decision).
    Or-nodes are added by some activation principles, as are all the edges in the feature-activation diagram.
    Table~\ref{tab:complexity:principles} shows the contribution of each (de)activation principle.
    It assumes the following values:
    \begin{itemize}
      \item $F$ -- the number of real-optional features in the feature model.
      \item $G < F$ -- the number of group features (either or or xor groups) in the feature model.
      \item $A_g < F$ -- the average size of group features (average number of features in a group).
      \item $X \leq G$ -- the number of xor-groups in the feature model.
      \item $A_x < F$ -- the average size of xor-groups (average number of features in an xor-group).
      \item $Ex$ -- the number of exclude constraints in the feature model.
      \item $R$ -- the number of requires constraints in the feature model.
    \end{itemize}    
    
    \begin{table*}[tbp]
      \centering
        \begin{tabular}{lllp{11cm}}
          \toprule
                          & Number of     & Number                                & \\ 
                          & or-nodes      & of edges                              & \\ 
                          & added         & added                                 & Comment\\ \midrule
          \actmand        &               & $F$                                   & Over-approximation: only mandatory children are activated, but there cannot be more than $F$ of those. \\
          \actpar         &               & $F$                                   & Over-approximation: not every feature is a parent, but there cannot be more than $F$ of those. \\
          \actreq         &               & $R$                                   & Over-approximation: we only add an edge for real-optional targets. \\
          \actgroup       & $G$           & $G\left(1+A_g\right)$                 & One or-node for each group with one edge into the or-node and one edge from the or-node to 
                                                                                    the feature decision for each feature in the group. \\
          \actxor         &               & $\left(X \cdot A_x \right) \cdot A_x$ & There are $X \cdot A_x$ features that are in an xor-group and each one of these needs an edge 
                                                                                    to each of it's neighbours---$A_x$ edges on average. \\
          \actexcl        &               & $Ex$                                  & Over-approximation: we only add an edge for real-optional targets. \\ \midrule
          \dechild        &               & $F$                                   & All features are child features of at most one parent feature, so at most one edge is added for them. \\
          \dexor\ / \deor & $G \cdot A_g$ & $G \cdot A_g\left(2+A_g\right)$       & Both principles do the same thing for or-groups and xor-groups, respectively. 
                                                                                    In total, they do this for all $G$ groups.
                                                                                    Add one or-node per feature in a group with 1 edge coming into the or-node, 1 edge from the or-node to the parent feature, and $A_g$ edges to all the sibling features. \\
          \deparent       &               & $F$                                   & Every feature has at most one parent so adds at most one edge as a result of this principle. \\
          \dereq          &               & $R$                                   & Over-approximation: we only add an edge for real-optional targets. \\
          \bottomrule
        \end{tabular}
      \caption{Contribution of (de)activation principles to the size of the feature-activation diagram.
              }
      \label{tab:complexity:principles}
    \end{table*}
    
    Overall, the size of the complete feature-activation diagram, and therefore the computational complexity of constructing it is
    \begin{align*}
        & O\left(\left(2F + G\left(1+A_g\right)\right) + \right.\\
        & \qquad \left. \left(4F + 2R + Ex + G + G \cdot A_g\left(3+A_g\right)+ X\cdot A_x^2\right)\right) \\
      =~ & O\left(F + R + Ex + G + G\cdot A_g^2 + X\cdot A_x^2\right)
    \end{align*}
    Given $G \leq F$ and assuming $C = R + Ex$ the number of cross-tree constraints, we can further simplify this to
    \[
    O\left(F + C + G\cdot A_g^2 + X\cdot A_x^2\right)
    \]
    
  \subsection{Creation of CPCOs}
  \label{section:appendix_complexity:cpco}
  
    CPCOs are created by depth-first search over the feature-activation sub-diagram starting at the root feature decision for the CPCO. 
    We construct $2F$ such CPCOs and, worst case, need to do a complete traversal of the feature-activation diagram for each one.
    After the traversal, we also need to resolve proxies for presence conditions and follow-ors.
    In the worst case, this requires touching the presence conditions of all feature decisions in the CPCO ($2F$ worst case).
    
    Thus, the computational complexity of constructing \emph{one} CPCO is
    \[
    O\left(F \cdot \left(F + C + G\cdot A_g^2 + X\cdot A_x^2\right)\right)
    \]
    and for \emph{all} CPCOs it is
    \[
    O\left(F \cdot F \cdot \left(F + C + G\cdot A_g^2 + X\cdot A_x^2\right)\right)
    \]
    
    One aspect that is not considered in the above complexity analysis is dead feature removal.
    This step applies a SAT solver on the full constraint generated to represent the VB-rule feature model. 
    This is used once for each feature in the VB-rule feature model (of which there are $O\left(G + G\cdot A_g\right)$).
    SAT is NP-complete, so even though modern SAT solvers are very efficient, in principle the dead-feature removal step can be very costly.
    It is worth noting that it is not an essential step, however: our CPCOs would work equally effectively if we did not make their representation more compact through the dead-feature removal step.
\section{Effect sizes}
\label{sec:appendix:effsize}
This appendix provides supplementary information for our discussion of results (Section~\ref{sec:results}).
Table~\ref{tab:EffectSizes} shows the observed effect sizes of our evaluation comparison by using the $A_{12}$ score (calculated using the R package \texttt{effsize}), following Vargha and Delaney's original interpretation~\cite{vargha2000critique}.

\begin{table}[t]
	\caption{Effect sizes in terms of $A_{12}$ \cite{vargha2000critique}.}
	\label{tab:EffectSizes}
	\centering
	\begin{threeparttable}
	\scriptsize
	\setlength{\tabcolsep}{0.5pt}
	\begin{tabular}{l|rr|rr}
		\toprule
		
		& \multicolumn{2}{c|}{\centering Result Quality (HV)}
		& \multicolumn{2}{c}{\centering Execution time (sec.)}
		\\
		Feature model
		& \makecell{\ACAPULCO\ vs. \\ \SATIBEA}
		& \makecell{\ACAPULCO\ vs. \\ \MODAGAME}
		& \makecell{\ACAPULCO\ vs. \\ \SATIBEA}
		& \makecell{\ACAPULCO\ vs. \\ \MODAGAME}
		\\

		\midrule
	    Wget          & 1                             & 0.504                          & 0.968                         & 0.068
	    \\
        Tank war       & 1                             & 0.957                           & 0.968                         & 0.893
        \\
        Mobile media   & 1                             & 0.869                           & 0.968                         & 0.948
        \\
        WeaFQAs        & 1                             & 1                               & 0.968                         & 0.968
        \\
        Busy Box       & 1                             & 1                               & 0.968                         & 1
        \\
        EMB ToolKit    & 1                             & 1                              & 0.968                         & 1
        \\
        CDL ea2468     & 1                             & 1                               & 0.322                         & 1
        \\
        Linux Distrib. & 1                             & 1                               & 0.968                         & 1
        \\
        Linux 2.6      & 1                             & -                            & 0.968                         & - 
        \\
        Automotive 2.1 & 1                             & -                            & 1                             & - 
        \\
		\bottomrule
	\end{tabular}
\begin{tablenotes}
    \item[] $A_{12}\!\approx\!0.56 = \textit{small}$; $A_{12}\!\approx\!0.64 = \textit{medium}$; and
	$A_{12}\!\gtrsim\!0.71 = \textit{large}$.
\end{tablenotes}
\end{threeparttable}
\end{table}

\section{Additional experiments}
\label{sec:appendix:add_exp}
This appendix provides further supplementary information for our discussion of results (Section~\ref{sec:results}), specifically, the results of additional experiments with modified termination criteria (20,000 insteaf of 5,000 evolutions).
Table~\ref{tab:AdditionalExperiments} gives an overview of the results for all feature models. 
Figures~\ref{fig:MobileMedia2FM_20000}, \ref{fig:WeaFQAsFM_20000}, and ~\ref{fig:linuxFM_20000} illustrate the results for the three representative feature models also chosen for illustration in Section~\ref{sec:results}.
Please note that the hypervolume scores shown here cannot be directly compared to those with 5,000 evolutions, because they were computed based on different reference Pareto fronts (see description in Sect.~\ref{sec:setup}).

\begin{figure}
  \centering
  \begin{subfigure}[t]{\PLOTSUBFIG}
    \centering
    \begin{tikzpicture}[scale=\PLOTWIDTH]
        \begin{axis}[
        xlabel={Evolutions},
        ylabel={Hypervolume},
        xmin=200, xmax=20000,
        ymin=0.0, ymax=0.6,
        legend pos=south east
        ]
        
        \addplot [\MDEOPLOTCOLOR,\MDEOPLOTSTYLE,\PLOTGENERALSTYLE] table [x=NFE, y={HV Median}, col sep=comma]{plotsdata/Acapulco_mobile_media2_statsResults_20000.dat};
        \addlegendentry{Acapulco}
        \addplot [\MODAGAMEPLOTCOLOR,\MODAGAMEPLOTSTYLE,\PLOTGENERALSTYLE] table [x=NFE, y={HV Median}, col sep=comma]{plotsdata/Modagame_mobile_media2_statsResults_20000.dat};
        \addlegendentry{Modagame}
        \addplot [\SATIBEAPLOTCOLOR,\SATIBEAPLOTSTYLE,\PLOTGENERALSTYLE] table [x=NFE, y={HV Median}, col sep=comma]{plotsdata/Satibea_mobile_media2_statsResults_20000.dat};
        \addlegendentry{Satibea}
        \end{axis}
    \end{tikzpicture}
    \vspace{-0.2cm}
    \caption{Hypervolume.}
    \label{sfig:MobileMedia2Vevols_20000}
  \end{subfigure}
  \begin{subfigure}[t]{\PLOTSUBFIG}
    \centering
    \begin{tikzpicture}[scale=\PLOTWIDTH]
        \begin{axis}[
        xlabel={Evolutions},
        ylabel={Execution time (s)},
        xmin=200, xmax=20000,
        ymin=0, ymax=2,
        legend pos=north west
        ]
        
        \addplot [\MDEOPLOTCOLOR,\MDEOPLOTSTYLE,\PLOTGENERALSTYLE] table [x=NFE, y={Time Median (s)}, col sep=comma]{plotsdata/Acapulco_mobile_media2_statsResults_20000.dat};
        \addlegendentry{Acapulco}
        \addplot [\MODAGAMEPLOTCOLOR,\MODAGAMEPLOTSTYLE,\PLOTGENERALSTYLE] table [x=NFE, y={Time Median (s)}, col sep=comma]{plotsdata/Modagame_mobile_media2_statsResults_20000.dat};
        \addlegendentry{Modagame}
         \addplot [\SATIBEAPLOTCOLOR,\SATIBEAPLOTSTYLE,\PLOTGENERALSTYLE] table [x=NFE, y={Time Median (s)}, col sep=comma]{plotsdata/Satibea_mobile_media2_statsResults_20000.dat};
        \addlegendentry{Satibea}
        \end{axis}
    \end{tikzpicture}
    \vspace{-0.2cm}
    \caption{Execution time.}
    \label{sfig:MobileMedia2ExecTime_20000}
  \end{subfigure}
   %
\vspace{-0.1cm}
  \caption{Additional experiments: results for Mobile Media.}
\vspace{0.3cm}
  \label{fig:MobileMedia2FM_20000}
  \begin{subfigure}[t]{\PLOTSUBFIG}
    \centering
    \begin{tikzpicture}[scale=\PLOTWIDTH]
        \begin{axis}[
        xlabel={Evolutions},
        ylabel={Hypervolume},
        xmin=200, xmax=20000,
        ymin=0, ymax=0.6,
        legend pos=north east
        ]
 
        \addplot [\MDEOPLOTCOLOR,\MDEOPLOTSTYLE,\PLOTGENERALSTYLE] table [x=NFE, y={HV Median}, col sep=comma]{plotsdata/Acapulco_weafqas_statsResults_20000.dat};
        \addlegendentry{Acapulco}
        \addplot [\MODAGAMEPLOTCOLOR,\MODAGAMEPLOTSTYLE,\PLOTGENERALSTYLE] table [x=NFE, y index={5}, col sep=comma]{plotsdata/Modagame_weafqas_statsResults_20000.dat};
        \addlegendentry{Modagame}
         \addplot [\SATIBEAPLOTCOLOR,\SATIBEAPLOTSTYLE,\PLOTGENERALSTYLE] table [x=NFE, y index={5}, col sep=comma]{plotsdata/Satibea_weafqas_statsResults_20000.dat};
        \addlegendentry{Satibea}
        \end{axis}
    \end{tikzpicture}
        \vspace{-0.2cm}
    \caption{Hypervolume.}
    \label{sfig:WeaFQAsHVevols_20000}
  \end{subfigure}
  \begin{subfigure}[t]{\PLOTSUBFIG}
    \centering
    \begin{tikzpicture}[scale=\PLOTWIDTH]
        \begin{axis}[
        xlabel={Evolutions},
        ylabel={Execution time (s)},
        xmin=200, xmax=20000,
        ymin=0, ymax=3,
        legend pos=north west
        ]
        
        \addplot [\MDEOPLOTCOLOR,\MDEOPLOTSTYLE,\PLOTGENERALSTYLE] table [x=NFE, y={Time Median (s)}, col sep=comma]{plotsdata/Acapulco_weafqas_statsResults_20000.dat};
        \addlegendentry{Acapulco}
        \addplot [\MODAGAMEPLOTCOLOR,\MODAGAMEPLOTSTYLE,\PLOTGENERALSTYLE] table [x=NFE, y={Time Median (s)}, col sep=comma]{plotsdata/Modagame_weafqas_statsResults_20000.dat};
        \addlegendentry{Modagame}
        \addplot [\SATIBEAPLOTCOLOR,\SATIBEAPLOTSTYLE,\PLOTGENERALSTYLE] table [x=NFE, y={Time Median (s)}, col sep=comma]{plotsdata/Satibea_weafqas_statsResults_20000.dat};
        \addlegendentry{Satibea}
        \end{axis}
    \end{tikzpicture}
        \vspace{-0.2cm}
    \caption{Execution time.}
    \label{sfig:WeaFQAsExecTime_20000}
  \end{subfigure}
   %
   \vspace{-0.1cm}
  \caption{Additional experiments: results for WeaFQAs.}\vspace{0.3cm}

  \label{fig:WeaFQAsFM_20000}
  \begin{subfigure}[t]{\PLOTSUBFIG}
    \centering
    \begin{tikzpicture}[scale=\PLOTWIDTH]
        \begin{axis}[
        xlabel={Evolutions},
        ylabel={Hypervolume},
        xmin=200, xmax=20000,
        ymin=0, ymax=0.6,
        legend pos=north west
        ]
        
        \addplot [\MDEOPLOTCOLOR,\MDEOPLOTSTYLE,\PLOTGENERALSTYLE] table [x=NFE, y={HV Median}, col sep=comma]{plotsdata/Acapulco_linux-2.6.33.3_statsResults_20000.dat};
        \addlegendentry{Acapulco}
         \addplot [\SATIBEAPLOTCOLOR,\SATIBEAPLOTSTYLE,\PLOTGENERALSTYLE] table [x=NFE, y={HV Median}, col sep=comma]{plotsdata/Satibea_linux-2.6.33.3_statsResults_20000.dat};
        \addlegendentry{Satibea}
        \end{axis}
    \end{tikzpicture}
        \vspace{-0.2cm}
    \caption{Hypervolume.}
    \label{sfig:linuxHVevols_20000}
  \end{subfigure}
  \begin{subfigure}[t]{\PLOTSUBFIG}
    \centering
    \begin{tikzpicture}[scale=\PLOTWIDTH]
        \begin{axis}[
        xlabel={Evolutions},
        ylabel={Execution time (s)},
        xmin=200, xmax=20000,
        ymin=0, ymax=10,
        legend pos=north west
        ]
        
        \addplot [\MDEOPLOTCOLOR,\MDEOPLOTSTYLE,\PLOTGENERALSTYLE] table [x=NFE, y={Time Median (s)}, col sep=comma]{plotsdata/Acapulco_linux-2.6.33.3_statsResults_20000.dat};
        \addlegendentry{Acapulco}
        \addplot [\SATIBEAPLOTCOLOR,\SATIBEAPLOTSTYLE,\PLOTGENERALSTYLE] table [x=NFE, y={Time Median (s)}, col sep=comma]{plotsdata/Satibea_linux-2.6.33.3_statsResults_20000.dat};
        \addlegendentry{Satibea}
        \end{axis}
    \end{tikzpicture}
        \vspace{-0.2cm}
    \caption{Execution time.}
    \label{sfig:linuxExecTime_20000}
  \end{subfigure}
   %
   \vspace{-0.1cm}
  \caption{Additional experiments: results for Linux 2.6.}
  \label{fig:linuxFM_20000}
\end{figure}
\begin{table*}[t]
	\caption{Additional experiments with termination criteria of 20,000 evolutions.}
	\label{tab:AdditionalExperiments}
	\centering
	\begin{threeparttable}
	\footnotesize
	\setlength{\tabcolsep}{1.5pt}
	\begin{tabular}{l|rrrr|rrrr|rrrrr||rr|rr}
		\toprule
		& 
		\multicolumn{4}{c|}{\centering \ACAPULCO}
		& \multicolumn{4}{c|}{\centering \MODAGAME}
		& \multicolumn{5}{c||}{\centering \SATIBEA}
		& \multicolumn{2}{c|}{\centering \ACAPULCO\ HV is greater}
		& \multicolumn{2}{c}{\centering \ACAPULCO\ time is faster}
		\\
		& \multicolumn{2}{c}{\centering HV}
		& \multicolumn{2}{c|}{\centering Time}
		& \multicolumn{2}{c}{\centering HV}
		& \multicolumn{2}{c|}{\centering Time}
		& \multicolumn{2}{c}{\centering HV}
		& \multicolumn{2}{c}{\centering Time}
		& \multicolumn{1}{c||}{\centering Invalid}
		
		& \multicolumn{1}{c}{\centering \MODAGAME}
		& \multicolumn{1}{c|}{\centering \SATIBEA}
		& \multicolumn{1}{c}{\centering \MODAGAME}
		& \multicolumn{1}{c}{\centering \SATIBEA}
		\\
		
		Feature model 
		& \multicolumn{1}{c}{\centering MD} & \multicolumn{1}{c}{\centering SD}
		& \multicolumn{1}{c}{\centering MD} & \multicolumn{1}{c|}{\centering SD}
		
		& \multicolumn{1}{c}{\centering MD} & \multicolumn{1}{c}{\centering SD}
		& \multicolumn{1}{c}{\centering MD} & \multicolumn{1}{c|}{\centering SD}
		
		& \multicolumn{1}{c}{\centering MD} & \multicolumn{1}{c}{\centering SD}
		& \multicolumn{1}{c}{\centering MD} & \multicolumn{1}{c}{\centering SD}
		& \multicolumn{1}{c||}{\centering Sols.}
		
		& \multicolumn{1}{c}{\centering $p$-value} 
		& \multicolumn{1}{c|}{\centering $p$-value} 
		
		& \multicolumn{1}{c}{\centering $p$-value} 
		& \multicolumn{1}{c}{\centering $p$-value} 
		\\
		
		\midrule

		Wget
		& \hlcell 0.44  & 3.52e-4
		& 0.92 & 0.08
		
		& 0.44 & 2.38e-4
		& \hlcell 0.88  & 0.06
		
		& 0.42 & 5.83e-3
		& 1.32  & 0.07
		& 2\%
		
		& 0.98
		& \hlcell 1.10e-11
		
		& 0.99
		& \hlcell 1.13e-10
		\\

		Tank war
		& \hlcell 0.46 & 9.04e-4
		& \hlcell 0.94  & 0.08

		& 0.46 & 1.06e-3
		& 1.01  & 0.06
		
		& 0.41 & 0.01
		& 1.38 & 0.07
		& 2\%
		
		& \hlcell 0.01
		& \hlcell 1.44e-11
		
		& \hlcell 2.66e-10
		& \hlcell 2.37e-11
		\\
	
		Mobile media
		& 0.48 & 2.40e-3
		& \hlcell 0.96 & 0.07

		& \hlcell 0.48 & 5.56e-4
		& 1.06 & 0.06
		
		& 0.42 & 0.02
		& 1.50 & 0.09
	    & 15\%
	    
	    & \hlcell 0.03
		& 0.99
		
		& \hlcell 2.42e-10
		& \hlcell 1.44e-11
		\\

		WeaFQAs
		& \hlcell 0.40 & 2.02e-3
		& \hlcell 1.15 & 0.08

		& 0.25 & 0.01
		& 2.06 & 0.07
		
		& 0.29 & 0.02 
		& 1.65 & 0.07
	    & 31\%
	    
	    & \hlcell 1.44e-11
        & \hlcell 1.44e-11
		
		& \hlcell 1.44e-11
        & \hlcell 1.44e-11
		\\
		
		
		
	
		
		
	    Busy Box
		& \hlcell 0.42 & 2.10e-3
		& \hlcell 1.31 & 0.10
		
		& 0.34 & 2.01e-3
		& 4.70 & 0.18
		
		& 0.35 & 4.31e-3
		& 2.25 & 0.09
	    & 24\%
	    
		& \hlcell 1.44e-11
		& \hlcell 1.44e-11
		
		& \hlcell 1.44e-11
		& \hlcell 1.44e-11
		\\
		
		EMB ToolKit
		& \hlcell 0.37 & 2.19e-3
		& \hlcell 2.30 & 0.11
		
		& 0.29 & 2.29e-3
		& 7.85 & 0.23
		
		& 0.32 & 4.46e-3
		& 3.87 & 0.11
	    & 62\%
	
		& \hlcell 1.44e-11
		& \hlcell 1.44e-11
		
		& \hlcell 1.44e-11
		& \hlcell 1.44e-11
		\\
		
		CDL ea2468
		& \hlcell 0.36 & 1.38e-3
		& \hlcell 3.17 & 0.13
		
		& 0.17 & 4.87e-3
		& 14.72 & 0.26
		
		& 0.30 & 6.91e-3
		& 3.42 & 0.10
		& 75\%
	
		& \hlcell 1.44e-11
		& \hlcell 1.44e-11
		
		& \hlcell 1.44e-11
		& \hlcell 4.66e-10
		\\
		
		Linux Distrib.
		& \hlcell 0.34 & 2.33e-3
		& \hlcell 0.52 & 0.07
		
		& 0.31 & 2.26e-3
		& 2.44 & 0.08
		
		& 0.30 & 1.17e-2
		& 0.65 & 0.07
	    & 37\%
	
		& \hlcell 1.44e-11
		& \hlcell 1.44e-11
		
		& \hlcell 1.44e-11
		& \hlcell 1.44e-11
		\\
		
		Linux 2.6
		& \hlcell 0.36 & 9.06e-4
		& \hlcell 5.66 & 0.18
		
		& - & -
		& - & -
		
		& 0.31 & 3.34e-3
		& 9.95 & 0.24
		& 89\%
	
		& -
		& \hlcell 1.44e-11
		
		& -
		& \hlcell 1.44e-11
		\\
		
		Automotive 2.1
		& \hlcell 0.34 & 4.20e-3
		& \hlcell 74.63 & 2.30

        & - & -
		& - & -
		
		& 0.01 & 5.17e-3
		& 151.73 & 5.71
	    & 98\%
	
        & -
		& \hlcell 1.44e-11
		
        & -
		& \hlcell 1.44e-11
		\\
	    
		\bottomrule
	\end{tabular}
\begin{tablenotes}
	\item [] Runs: 30. Population: 100. Generations: 200 (20,000 evolutions). 
\end{tablenotes}
\end{threeparttable}
 \vspace{-0.4cm}
\end{table*}





\end{document}